\documentclass[structabstract]{aa}  
\usepackage{graphicx}
\usepackage{dcolumn}
\usepackage{txfonts}
\usepackage{natbib}
\usepackage{xcolor}
\usepackage{color}
\usepackage{natbib,twoopt}
\bibpunct{(}{)}{;}{a}{}{,} 
\newcommandtwoopt{\citeads}[3][][]{\href{http://adsabs.harvard.edu/abs/#3}%
{\citealp[#1][#2]{#3}}}
\newcommandtwoopt{\citepads}[3][][]{\href{http://adsabs.harvard.edu/abs/#3}%
{\citep[#1][#2]{#3}}}
\newcommandtwoopt{\citetads}[3][][]{\href{http://adsabs.harvard.edu/abs/#3}%
{\citet[#1][#2]{#3}}} 
\newcommandtwoopt{\citeyearads}[3][][]%
{\href{http://adsabs.harvard.edu/abs/#3}{\citeyear[#1][#2]{#3}}}

\newcommand\kep{\emph{Kepler}}

\begin{document}
\authorrunning{S. Mathur et al.}
\title{Detections of solar-like oscillations in dwarfs and subgiants with \kep\, DR25 short-cadence data}
   \titlerunning{Detections of solar-like oscillations in \kep\, DR25 SC data}


  \author{S. Mathur\inst{1,2}
          \and
          R.~A. Garc\'ia\inst{3} 
                      \and
            S. Breton\inst{3}
          \and
          A. R. G. Santos\inst{4,5}
            \and
          B. Mosser\inst{6}
            \and
            D. Huber\inst{7}
          \and
          M. Sayeed\inst{7,8}
            \and
          L. Bugnet\inst{9,3} 
          \and
          A. Chontos\inst{7,10}
          }
\institute{Instituto de Astrof\'isica de Canarias (IAC), E-38205 La Laguna, Tenerife, Spain\\ \email{smathur@iac.es}
\and
Universidad de La Laguna (ULL), Departamento de Astrof\'isica, E-38206 La Laguna, Tenerife, Spain 
\and
AIM, CEA, CNRS, Universit\'e Paris-Saclay, Universit\'e de Paris, Sorbonne Paris Cit\'e, F-91191 Gif-sur-Yvette, France
\and
Department of Physics, University of Warwick, Coventry, CV4 7AL, UK
\and
Space Science Institute, 4765 Walnut Street, Suite B, Boulder CO 80301, USA
\and
LESIA, Observatoire de Paris, Universit\'e PSL, CNRS, Sorbonne Universit\'e, Universit\'e de Paris, 92195 Meudon, France
\and
Institute for Astronomy, University of Hawai`i, 2680 Woodlawn Drive, Honolulu, HI 96822, USA
\and
Department of Astronomy, Columbia University, 550 West 120th Street, New York, NY, USA 
\and
Flatiron Institute, Simons Foundation, 162 Fifth Ave, New York, NY 10010, USA
\and
NSF Graduate Research Fellow
}
  
  \date{Received August 18, 2011; accepted }

 
  \abstract
    {During the survey phase of the \kep\ mission, several thousands of stars were observed in short cadence, allowing the detection of solar-like oscillations in more than 500 main-sequence and sub-giant stars. These detections showed the power of asteroseismology to determine the stellar fundamental parameters. However, the \kep\ Science Office discovered an issue in the calibration that affected half of the short-cadence data, leading to a new data release (DR25) with improved corrections of the lightcurves. We re-analyze here the one-month time series of the \kep\ survey phase to search for solar-like oscillations that might have been missed when using the previous data release. We study the seismic parameters of 99 stars among which 46 targets with new reported solar-like oscillations, increasing by around 8\% the known sample of solar-like stars with asteroseismic analysis {  of the short-cadence data} from this mission. The majority of these stars have mid- to high-resolution spectroscopy  publicly available with the LAMOST and APOGEE surveys respectively as well as precise {\it Gaia} parallaxes. We compute the masses and radii using seismic scaling relations and find that this new sample populates the massive stars (above 1.2\,$M_\odot$ {  and up to 2\,$M_{\odot}$}) and subgiant phase. We determine the granulation parameters and amplitude of the modes, which agree with the scaling relations derived for dwarfs and subgiants. The stars studied here are slightly fainter than the previously known sample of main-sequence and subgiants with asteroseismic detections. We also study the surface rotation and magnetic activity levels of those stars. {  Our sample of 99 stars has} similar levels of activity compared to the previously known sample and in the same range as the Sun between the minimum and maximum of its activity cycle. {  We find that for seven stars, a possible blend could be the reason for the non detection with an early data release.} Finally we compare the radii obtained from the scaling relations with the {\it Gaia} ones and find that the {\it Gaia} radii are overestimated by {\color {black} 4.4\%} {  on} average compared to the seismic radii {  with a scatter of 12.3\% and}  a decreasing trend with evolutionary stage. {   In addition, for homogeneity purposes, we re-analyze the DR25 of the main-sequence and sub-giant stars with solar-like oscillations previously detected and provide their global seismic parameters for a total of 526 stars.} }


   \maketitle
%

\section{Introduction}

The NASA {\it Kepler} main mission has shown the power of using asteroseismology to precisely characterize the properties of solar-like stars \citep[e.g.][]{2017ApJ...835..173S,2017ApJS..233...23S}. Based on the latest catalogs of \kep\ stellar properties \citep{2014ApJS..211....2H,2017ApJS..229...30M,2018ApJ...866...99B,2020AJ....159..280B}, more than 125,000 stars are supposed to be main-sequence solar-like stars according to their effective temperatures, surface gravities as well as the {\it Gaia} DR2 observations \citep[e.g.][]{2018A&A...616A...8A,2018A&A...616A...1G}, which were included in the latest \kep\ catalog. The turbulence of the outer layers of those stars excite oscillations known as solar-like oscillations \citep[e.g.][]{2019LRSP...16....4G}. Given the radius, mass, and gravity of these stars, the modes are expected to be above 300\,$\mu$Hz, which means that we need to observe them in short cadence (sampling of $\sim$\,58.85\,s) in order to be able to detect the acoustic modes. 

\noindent During the survey phase of the \kep\ mission, around 2,600 stars were observed between May 2009 and March 2010 for a duration of approximately one month each and in a short-cadence mode. \citet{2011Sci...332..213C} performed a seismic analysis of the data available at that time and were able to detect solar-like oscillations in more than 500 stars. {\color {black} This led to the seismic characterization of that sample, providing their masses, radii, and ages \citep{2017ApJS..233...23S}. More recently, \citet[][]{2020FrASS...7...85B} analyzed hot main-sequence stars with $T_{\rm eff} >$\,6,000\,K in order to study the location of the $\delta$ Scuti/$\gamma$ Doradus instability strip and detected solar-like oscillations in 70 new \kep\, targets.}

\noindent Later, for the close-out of the nominal \kep\ mission in 2016, a new data release \citep[DR25,][]{Thompson2016} \footnote{\url{http://archive.stsci.edu/kepler/release_notes/release_notes25/KSCI-19065-002DRN25.pdf}} was done by the \kep\ Science Office. The calibration of the data was improved, in particular how some instrumental effects were corrected, which had an impact on about 50\% of the short-cadence data.
 

\noindent While the effect of the calibration should be negligible, the \kep\ Science Office could not assess the impact on the asteroseismic studies.  Using the newly calibrated data of DR25, \citet{2017EPJWC.16001007S} re-analyzed the new lightcurves for 18 solar-analogs. In particular, the authors re-computed their global seismic parameters such as the mean large frequency separation that scales with the root-mean-square of the mean stellar density \citep{1995A&A...293...87K} and the frequency of maximum oscillation power that is related to the surface gravity of the star \citep{1991ApJ...371..396B}. Comparing the new results with those obtained with the previous data release ({  DR23}), \citet{2017EPJWC.16001007S} found that the impact on the measurement of the mean large frequency separation, the frequency of maximum oscillation power, and the background parameters was negligible for the stars with high signal-to-noise ratio (SNR). {  Nevertheless,} the improvement of the lightcurves could still have an impact on the detection of the modes. With this in mind, we conduct the current work to determine the impact of the improved DR25 on the mode detection for \kep\ solar-like stars.

{  \noindent One would hope that the newly calibrated DR25 data would lead to a higher number of detections of solar-like oscillations. However, that is without taking into account the surface magnetism of stars. Indeed, for the Sun and several solar-like stars, it has been shown that the amplitude of the modes is anti-correlated with magnetic activity \citep[e.g.][]{2010Sci...329.1032G,2015MNRAS.454.4120H,2017A&A...598A..77K,2018ApJS..237...17S}. As a consequence, the non detection of modes in the remaining stars can partially be explained by the high magnetic activity level of the stars \citep{2011ApJ...732L...5C} but other factors, such as metallicity or binarity, can also be at play \citep{2019FrASS...6...46M}.}

\noindent {  In order to learn about the magnetic activity of stars with photometric data, we can use rotational modulation that results from the presence of active regions on the stellar surface. Additionally, the study of stellar rotation can provide key information for the understanding of angular momentum transport {  \citep[e.g.][]{2019ARA&A..57...35A,2019ApJ...872..128V,2020AJ....160...90A,2020ApJ...904..140C,2021ApJ...912..127S}}. Several studies have been led with \kep\ and K2 data to measure surface rotation periods in a large sample of stars \citep[e.g.][]{2014ApJS..211...24M,2019ApJS..244...21S,2020A&A...635A..43R,2021ApJ...913...70G,2021ApJS..255...17S}. Furthermore, combined with precise ages of stars, such as those obtained with asteroseismology or for clusters, it is possible to improve rotation-age relationships \citep[e.g.][]{2003ApJ...586..464B,2008ApJ...687.1264M,2015MNRAS.450.1787A,2016Natur.529..181V,2021NatAs.tmp...71H,2021arXiv210101183G}.}

\noindent In Section 2, we describe the data that are used here and the procedure to calibrate the lightcurves to optimize them for asteroseismic studies. Then we explain how the search for the acoustic modes was done in Section 3 and consolidate the sample along with the stellar atmospheric parameters. Section 4 presents the study of the convective background parameters as well as the maximum amplitude of the modes and their correlation with the global seismic parameters.  In Section 5, we compute the masses and radii based on the seismic scaling relations and look at the rotation and magnetism of this new sample compared to the previously known sample of solar-like stars with detections of acoustic modes. Finally, we finish with the discussion on the stellar parameters of the new seismic detections and provide a summary of this work (Sections 6 and 7).

\begin{figure*}[!h]
\begin{center}
\includegraphics[width=15cm]{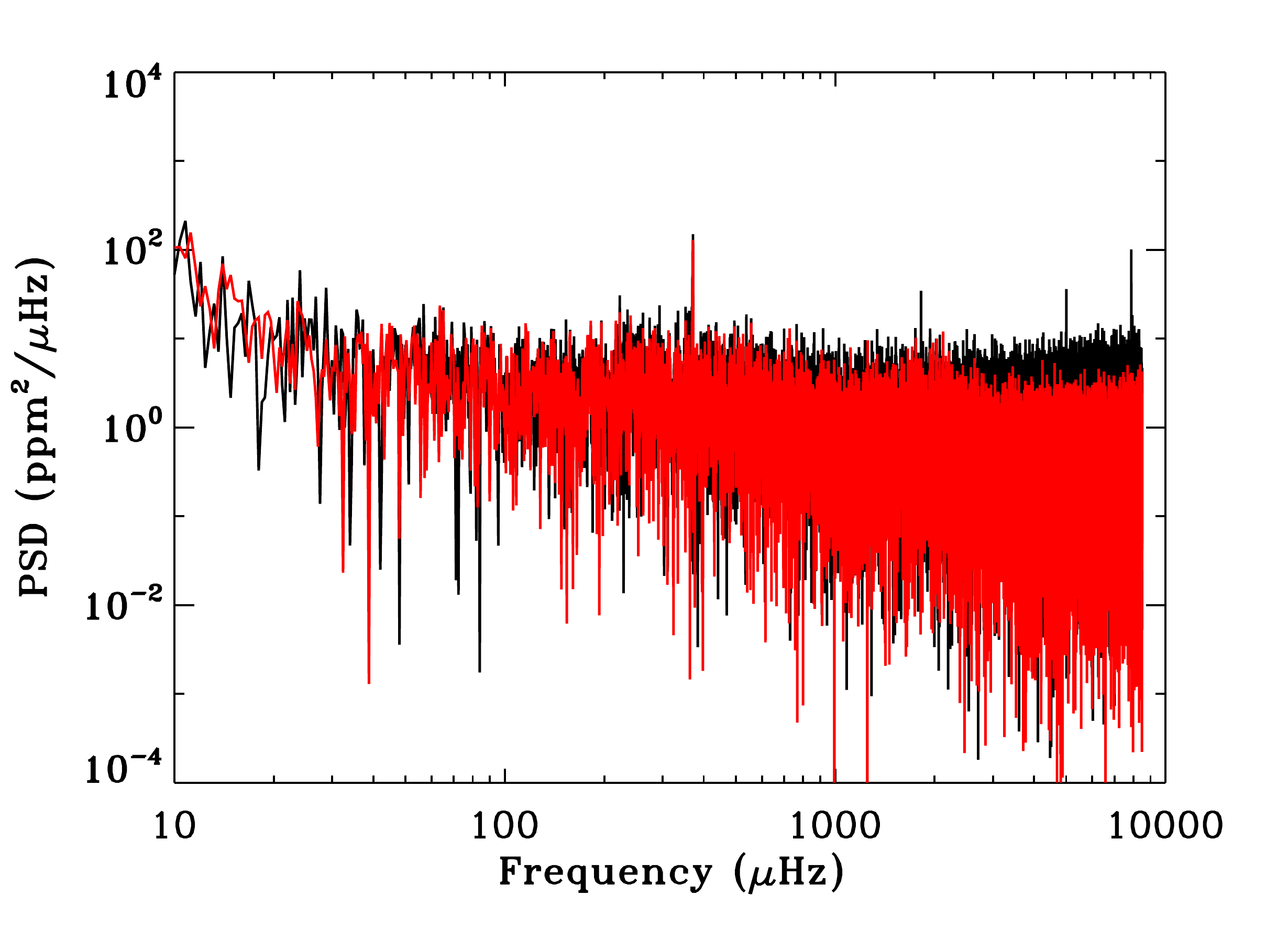}
\caption{Comparison of the Power Spectrum Density of KIC~4484238 using the DR24 data (black curve) and the DR25 data (red curve).}
\label{PSD_DR24_25}
\end{center}
\end{figure*}

\section{Data description}\label{sec:Data}


Here we processed both the short-cadence (SC) data (sampling of 58.85s) and the long-cadence (LC) data (sampling of $\sim$\,29.42min). The SC data were used for the asteroseismic analysis whereas the LC data were used for the study of the rotation and magnetic activity as for those phenomena a high cadence is not necessary. 

\noindent For the SC, the lightcurves used in this analysis are based on the last \kep\ data release DR25, where the \kep\ Science Office improved the calibration of the data. Indeed, as mentioned earlier, since the beginning of the mission, the short-cadence lightcurves were produced by correcting for the effects of {\it Kepler} shutterless readout. This scrambled the collateral smear and affected half of the short-cadence data.  Hence, Presearch Data Conditioning - Simple Aperture Photometry \citep[PDC-SAP;][]{2010ApJ...713L..87J} lightcurves were obtained from the Mikulski Archive for Space Telescopes (MAST)\footnote{\url {https://mast.stsci.edu/portal/Mashup/Clients/Mast/Portal.html}}.

\noindent The LC data were also obtained from the MAST but the calibrated lightcurves produced by the \kep\ Science Office are not necessarily optimized for rotation studies as some filters are applied. {  Therefore, from the target pixel files, we first selected a larger aperture. Starting by the pixel of the center of the star, adjacent pixels are added to the aperture if the average flux of the new pixel is decreasing (i.e., new added pixels should have less flux than the previous one in each direction). The addition of pixels to the aperture is stopped if: a) the mean flux of the new pixel increases, which suggests the presence of a nearby star; or b) the mean flux of the new pixel is less than a given threshold, which is established at 100 e-/s. This threshold allows us to maximize the flux from the target and yields larger apertures than those employed for the standard PDC-SAP ones.}




\noindent For both cadences, the obtained lightcurves were then calibrated for asteroseismology with the \kep\ Asteroseismic Data Analysis Calibration Software \citep[KADACS;][]{2011MNRAS.414L...6G}. We first removed outliers, corrected jumps, and removed instrumental trends. We also filled the gaps using the inpainting technique (with a multiscale discrete cosine transform) as described in \citet{2014A&A...568A..10G} and \citet{2015A&A...574A..18P}. This gap-filling technique allows us to reduce the noise at all frequencies, especially for lightcurves with rotation modulation. For the long-cadence lighcurves, we concatenated the different quarters and applied three high-pass filters with a cut-off period of 20, 55, and 80 days. Both short-cadence and long-cadence corrected lightcurves, called KEPSEISMIC, are available on the MAST\footnote{\url {https://archive.stsci.edu/prepds/kepseismic/}}.

\noindent In Figure~\ref{PSD_DR24_25}, we show the example of KIC~4484238, where the DR25 lightcurve leads to a significant improvement in the SNR. Indeed, while with the previous data release we could not detected any excess of power due to the oscillation modes, the DR25 power spectrum density (PSD) presents a bump around 2000\, $\mu$Hz.

\noindent Finally, for the long-cadence data, we also used the PDC-MAP \citep[Maximum A Posteriori,][]{Thompson_2013} data to which we applied the same KADACS correction software and gap filling technique to check the rotation period retrieved from the lightcurves and to make sure that there is no nearby star polluting the aperture.\\







\section{Looking for new detection of solar-like oscillations}\label{sec:detection}

\noindent Using the short-cadence lightcurves calibrated in Section~\ref{sec:Data}, we first performed a blind analysis of 2,572 lighcurves with the A2Z pipeline \citep{2010A&A...511A..46M}. This analysis provided global seismic parameters, in particular the mean large frequency separation, $\Delta \nu$, and the frequency of maximum power, $\nu_{\rm max}$. We then visually checked the stars for which the relation between $\Delta \nu$ and $\nu_{\rm max}$ agreed {  within 20\% of the empirical relation derived from the {\it Kepler} data \citep{2011ApJ...743..143H}}. This led to a sample of 20 stars where the detection seemed to be clear. {  In addition, following the procedure of \citet{2010A&A...511A..46M}, we found that the measurement of $\Delta \nu$ had} more than 95\% probability of being due to stellar signal. 

\noindent We then visually checked the outputs of the A2Z pipeline as well as the power spectrum densities of the remaining stars, flagging some clear detections as well as some candidates. We also computed the FliPer metric \citep[Flicker in Power,][]{2018A&A...620A..38B}. {  We used the sklearn.ensemble.RandomForest Regressor function \citep{2012arXiv1201.0490P} to train and combine predictions from a set of $200$ independent random trees. By combining the effective temperature of the star (\citet{2014ApJS..211....2H} and \citet[][hereafter M17]{2017ApJS..229...30M}) and FliPer granulation parameters \citep[see][for more details about the intrinsic methodology and the training of the supervised machine learning algorithm]{2018A&A...620A..38B}, the trained Random Forest \citep{2001MachL..45....5B} provides rough estimates of the frequency of maximum power. Based on the power contained in the power spectrum, this procedure flags stars that have a power spectrum consistent with the ones of solar-like stars.} This allowed us to select 20 additional stars where the blind run of A2Z agreed within 20\% of the FliPer predicted $\nu_{\rm max}$.






{\noindent For the remaining stars, we ran again the A2Z pipeline forcing $\nu_{\rm max}$ to the predicted value from the FliPer metric. For that subsample, we selected the stars where $\Delta \nu$ agreed within 20\% with the value expected from $\nu_{\rm max}$ based on the seismic scaling relations.

\noindent These steps led to a total sample of {\color {black} 105} stars with possible or confirmed detections. {\color {black} After cross-checking with the sample of new detections by  \citet[][hereafter Ba20]{2020FrASS...7...85B}, we added 43 stars to the analysis that were not selected and were part of our original sample. Our sample of 148 targets includes 60 stars that are part of the new detections of Ba20.} 

\noindent All the stars were also analyzed independently by the COR pipeline \citep{2009A&A...508..877M}, in a blind way first, and then in an iterative way consisting in using the seismic information provided by the $\nu_{\rm max}$ values derived by A2Z, in order to check for a possible agreement with this pipeline. We furthermore analyzed the sample with pySYD\footnote{https://github.com/ashleychontos/pySYD} {  \citep{2021arXiv210800582C}}, an open-source python-based adaptation of the SYD pipeline \citep{2009CoAst.160...74H}.



\subsection{Consolidating surface gravity, effective temperature, and metallicity from literature}\label{sec:atmosph}

 In order to better characterize the stellar parameters, we consolidated the atmospheric parameters for the full sample of {\color {black}148} stars by collecting values available in the literature. While all targets had $\log g$, $T_{\rm eff}$, and [Fe/H] available from the M17, many of those values were obtained from photometry, which could have large uncertainties. However, since then a few spectroscopic surveys have observed many of the targets in the \kep\ field. 

\noindent {  This is the case of the APOGEE survey \citep[Apache Point Observatory for Galactic Evolution Experiment,][]{gunn+2006,2019PASP..131e5001W}, which is part of the SDSS-IV \citep[Sloan Digital Sky Survey-IV][]{2017AJ....154...28B} and that provided high-resolution spectroscopy for more than 15,000 \kep\, targets. We used the DR16 \citep{2020ApJS..249....3A} for which the data were calibrated following \citet{2015AJ....150..148H}, \citet{2015AJ....150..173N}, and \citet{2016AJ....151..144G}. The majority of those stars} were red giants {  and a few hundreds were dwarfs and subgiants as part of  APOKASC \citep{2017ApJS..233...23S,2018ApJS..239...32P}}, a collaboration between APOGEE and the \kep\, Asteroseismic Science Consortium, but they also included several thousands of dwarfs without seismic detections for other science objectives.   

\noindent Another large spectroscopic survey, the Large Sky Area Multi-ObjectFiber Spectroscopic Telescope \citep[LAMOST,][]{2012RAA....12..723Z,2015ApJS..220...19D,2018ApJS..238...30Z}, also observed the \kep\, field with low-resolution spectroscopy.

\noindent Finally, the \emph{Gaia} mission is also providing precise parallaxes for a large sample of the \kep\, field stars.  \citet[][hereafter B20]{2020AJ....159..280B} derived new stellar parameters for the \kep\,-{\it Gaia} targets by fitting isochrones using the \emph{Gaia} DR2 {  \citep{2018A&A...616A...1G}.}

\noindent We used the following prioritization to build our list of spectroscopic parameters for the {\color {black} 148} targets previously selected. When available we first selected the stellar parameters obtained with high-resolution spectroscopy from APOGEE. Then we completed the parameters with low-resolution spectroscopy from LAMOST DR5 \citep{2018MNRAS.477.4641R}. Finally we took the stellar parameters from the latest \kep\,-{\emph{Gaia} stellar parameters catalog.  {\color {black} We found 102 stars with APOGEE spectra, 33 stars with LAMOST stellar parameters, and 13 stars with \emph{Gaia} parameters.}

\subsection{Finalizing the sample of confirmed detections}\label{sec:final}

The comparison of the results obtained by the three seismic pipelines, A2Z, COR, and pySYD allowed us to build a list of confirmed detections constituted of {\color {black} 99} stars where at least two pipelines agreed within 10\% in $\nu_{\rm max}$ as it has already been demonstrated that such scatter between pipelines is reasonable in particular for low SNR cases \citep[e.g.][]{2020ApJS..251...23Z}. In the remainder of the paper, we will refer to those 99 stars as {\it our sample}.

\noindent The final step was to consolidate the values of the mean large frequency separation, $\Delta \nu$, which was more difficult to determine for the cases with very low SNR and low resolution. We visually checked the \'echelle diagrams of the {\color {black} 99} targets and found 19 stars where the value of $\Delta \nu$ could be improved to have straighter ridges for the radial modes. So we refined $\Delta \nu$ by applying an additional analysis with the A2Z+ pipeline where we created a template to mimic the modes with five orders where the initial value for $\Delta \nu$  was taken as the value expected from scaling relations. We cross-correlated that template with the region of the PSD around $\nu_{\rm max}$ and swept the value of the mean large separation by +/-20\%. The value for which we obtained the maximum correlation is taken as the new $\Delta \nu$. This allowed us to converge for 7 stars. For the remaining 12 stars (KIC~3124465, 5818478, 6289367, 7009852, 7255919, 7598321, 7708535, 8349736, 8652398, 9892947, 9894195, 9912680), we selected $\Delta \nu$ that straightened the ridges. The associated uncertainties for those stars were obtained by computing the difference between the selected value and the value obtained by the A2Z pipeline. 


\noindent For the 60 stars in common with the sample of Ba20, we confirmed the seismic detections for {  53} stars. The values of $\nu_{\rm max}$ of these stars are in agreement with the values reported by Ba20 within 10\%. {  More specifically, there is an average offset of 4.3\% in $\nu_{\rm max}$ with a dispersion of 9.2\% and an average offset of 1.4\% in $\Delta \nu$ with a dispersion of 4.1\%.} The mean large frequency spacings agree within 5\% except for KIC~7669332, which appeared as an outlier in Figure 3 of Ba20. Our visual check of the \'echelle diagram confirms the reliability of our estimation. We also note that two targets of our sample (KIC~7215603 and 9715099) were included in \citet{2013ApJ...767..127H} and \citet[][hereafter C14]{2014ApJS..210....1C} but no $\nu_{\rm max}$ values were reported due to the aforementioned issue in the DR24.

\noindent We tested whether the apparent large fraction of stars more massive than usually observed in previous seismic data sets could be related to any artifact due to, for instance, the low signal-to-noise of the detection. Any effect translating in an overestimation of $\nu_{\rm max}$ or underestimation of $\Delta \nu$ will translate in an overestimated stellar mass derived from the seismic scaling relation. A general bias toward large $\nu_{\rm max}$ value is not expected for seismic data dominated by noise \citep[Eq. 23 of][]{2019A&A...622A..76M}. In order to identify any bias in $\Delta \nu$, we used the formalism developed by \citet{2013A&A...550A.126M}, which expresses the oscillation pattern in a parametric form. Therefore, we slightly smoothed the oscillation spectrum, with a smoothing function with a typical width of $\Delta \nu / 50$, and correlated it with the generic oscillation pattern defined by \citet{2013A&A...550A.126M}. This allowed us to do a more precise measurement of the large separation, so that we can consider that the fraction of stars more massive than usually observed is real.


\noindent In Table~\ref{tab1} we provide the global seismic parameters as well as the atmospheric parameters of the {\color {black} 99 stars with confirmed detection of solar-like oscillations}.



\begin{figure}[htbp]
\begin{center}
\includegraphics[width=9cm]{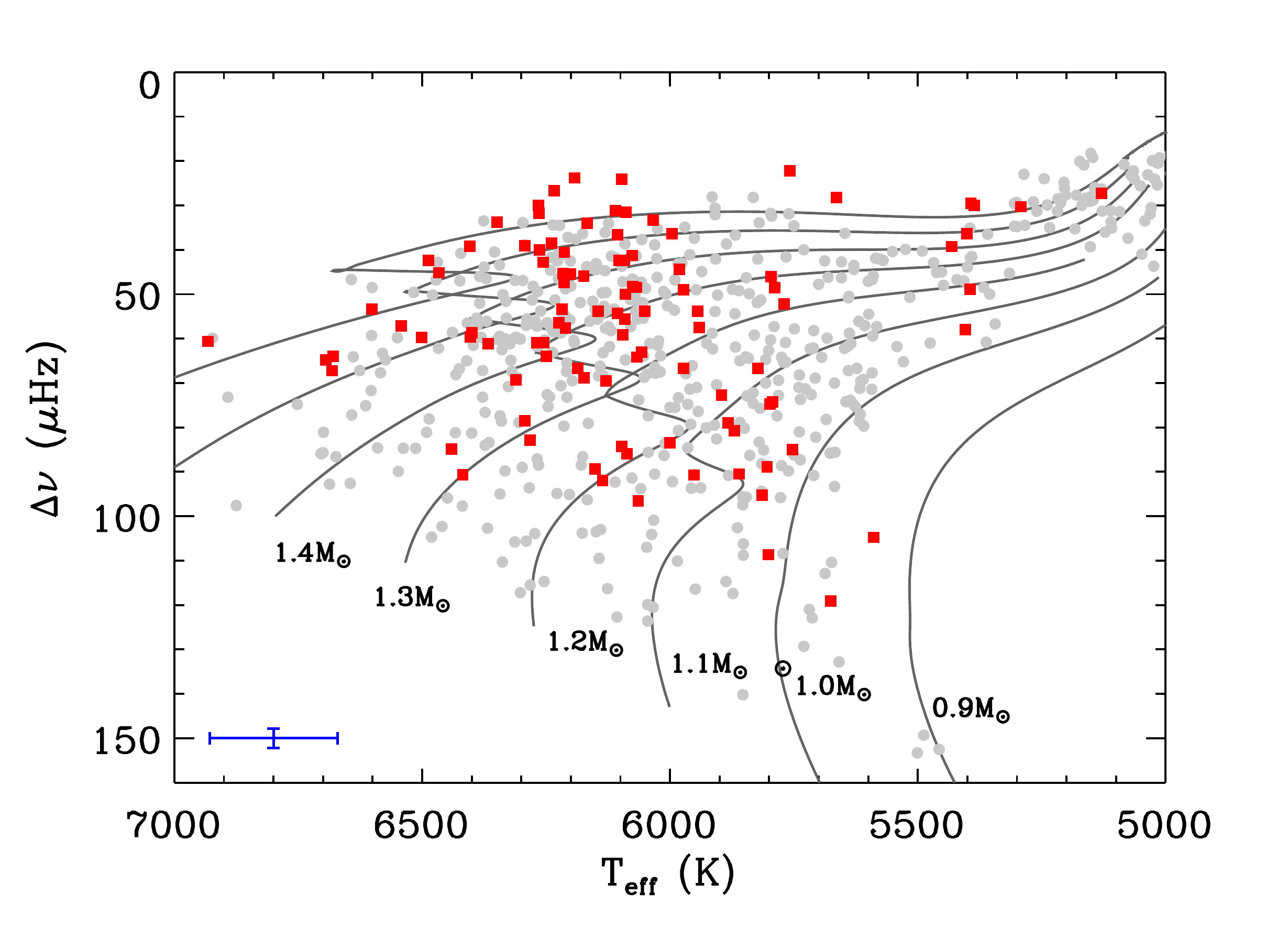}
\caption{Seismic Hertzsprung-Russell Diagram where the mean large frequency separation is used instead of the luminosity. The C14 solar-like stars are represented with grey circles where the effective temperature is taken from the {  M17}. The {\color {black} 99} stars with confirmed seismic detections are shown with red squares where the effective temperature is coming from APOGEE, LAMOST or B20 (see Section~\ref{sec:atmosph}). The position of the Sun is indicated by the $\odot$ symbol and the grey solid lines represent evolution tracks from ASTEC {  \citep{2008Ap&SS.316...13C}} for a range of masses at solar composition (Z$_\odot$ = 0.0246). Typical uncertainties are represented in the bottom left corner.}
\label{HRD}
\end{center}
\end{figure}

\noindent Figure~\ref{HRD} represents the seismic Hertzsprung-Russell Diagram of the \kep\ solar-like stars for which C14 had detected oscillations (grey symbols) along with the new confirmed detections (red symbols). We can see that many of the {  stars in our sample} are hotter than the Sun, hence more on the massive side, as well as more evolved solar-like stars, with a large fraction of subgiants. We also populate more the early red giant region where C14 had reported around tens of such evolved stars. 

\noindent {  We note that six of our targets have a flag 'BinDet\_NoCorr' in the B20 catalog, suggesting that they are binary candidates, which will have an impact on the derived effective temperature towards the red. We then looked for the {\it Gaia} DR2 effective temperatures of those stars. For four of them the values agreed with the ones reported by the spectroscopic surveys within the uncertainties. For one star, there is no {\it Gaia} DR2 $T_{\rm eff}$. For the target KIC~3633538, the {\it Gaia} effective temperature is of 5676\,K compared to 5100\,K in B20. The lower temperature in B20 can be explained by the binarity so we use the {\it Gaia} effective temperature for that star. This higher value is also more compatible with the location of the solar-like oscillation modes for that star.}


\begin{figure}[htbp]
\begin{center}
\includegraphics[width=9cm]{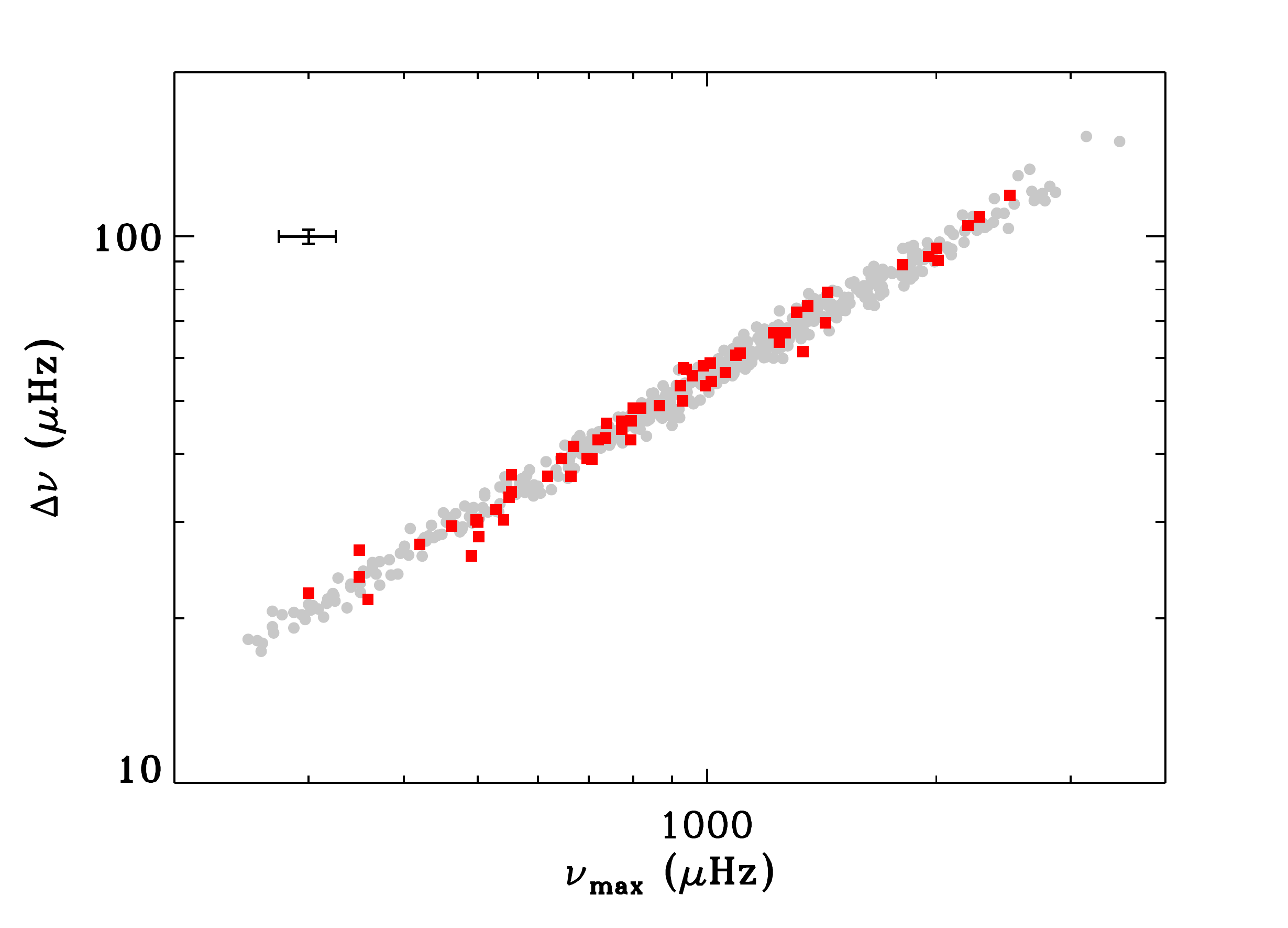}
\caption{Mean large frequency spacing, $\Delta \nu$, as a function of the frequency of maximum power, $\nu_{\rm max}$ for the C14 sample (grey circles) and {  our sample} (red squares). Typical uncertainties are shown in the upper left corner.}
\label{dnu_numax}
\end{center}
\end{figure}

\noindent The results of the global seismic analysis are shown in a $\Delta \nu$-$\nu_{\rm max}$ diagram (Figure~\ref{dnu_numax}) for the {  stars in our sample} (red squares) to be compared to the C14 sample (grey circles). We can see that the relations are very similar. Nevertheless, we can also note some of the {  stars in our sample} below the general trends, {  in particular near $\nu_{\rm max}$ of 500\,$\mu$Hz}, suggesting that those stars are slightly more massive than the rest of the sample, which was shown for instance by \cite{2010A&A...517A..22M}. The average uncertainties are of 2.4\% in $\nu_{\rm max}$ and 5.8\% on $\Delta \nu$. The latter is slightly high compared to the usually 5\% reported in C14 but representative of the low SNR.

\noindent Using the surface gravity and effective temperature in the seismic scaling relations, we can estimate a predicted value for the frequency of the maximum power as follows:


\begin{equation}
\nu_{\rm max} = \nu_{\rm max, \odot} \frac{g}{g_{\odot} \sqrt{T_{\rm eff}/T_{\rm eff, \odot}}},
\end{equation}

\noindent where $\nu_{\rm max, \odot}$ is the frequency of maximum power for the Sun, taken as 3100\,$\mu$Hz, $T_{\rm eff, \odot}$= 5777\,K, and $\log g_{\odot}$=4.4377\,dex.

\noindent In Figure~\ref{numax_comp}, the measured $\nu_{\rm max}$ for the confirmed sample is compared to the predicted value, $\nu_{\rm max, pred}$ from Eq. 1. Here we used the $\log g$ from B20 as they have been shown to be more reliable. The B20 catalog contains 186,301 \kep\ stars and a surface gravity value was available for {\color {black} 97} of our stars. The agreement between the observed and the predicted $\nu_{\rm max}$ is in general within 20\% with an average discrepancy of {\color {black} 1.2\%}, with the predicted $\nu_{\rm max}$ overestimating the observed one. However the disagreement can reach up to 50\%. 

\begin{figure}[h!]
\begin{center}
\includegraphics[width=9cm]{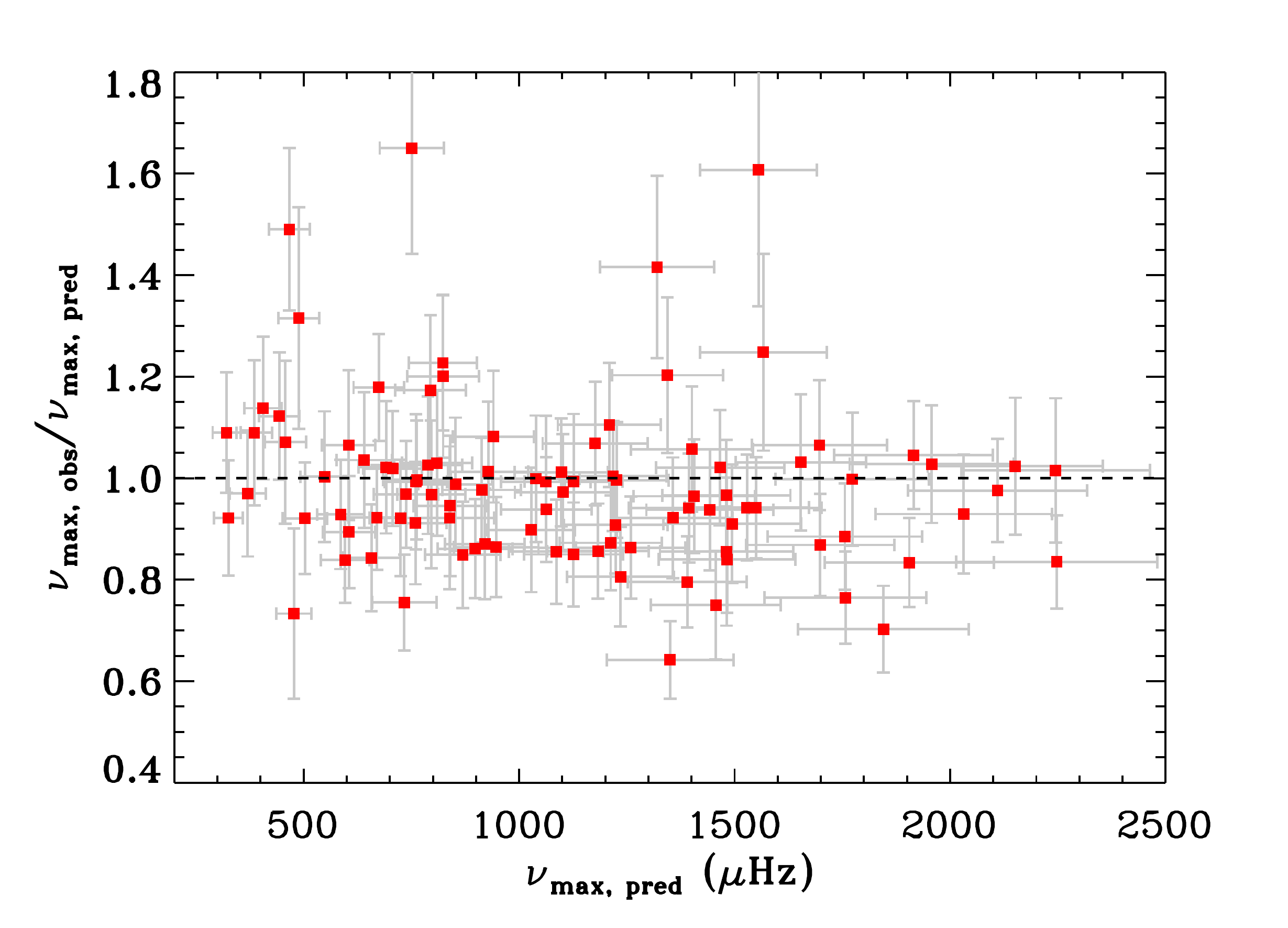}
\caption{Ratio of the observed $\nu_{\rm max, obs}$ and the predicted $\nu_{\rm max, pred}$ using the $\log g$ from B20. The {  dashed} line shows the equality of both values.}
\label{numax_comp}
\end{center}
\end{figure}

\noindent We did a similar comparison by using the spectroscopic surface gravities (see Appendix A) and the disagreement is larger with an average discrepancy of {\color {black} 30\%}, again with an overestimation of the prediction.


\noindent We note that around 10 stars have high enough SNR to fit the individual modes. The determination of the frequencies of the individual modes will allow us to do boutique modeling for that subsample of stars, which will provide more precise stellar parameters as well as ages. However this is out of the scope of this paper and is part of a subsequent paper (Mathur et al. in prep.).

\noindent We also visually checked the \'echelle diagrams of the remaining 49 stars of the 148 stars without confirmation of acoustic-mode detection. A list of candidates of 26 stars was kept where either the \'echelle diagram seemed to show some ridges, though with a very poor SNR, or when there was some agreement with the predicted $\nu_{\rm max}$. Note that some of these candidates did not necessarily have an agreement between pipelines and three of them are part of the Ba20 sample. We show the properties of those candidates in the Appendix B.

\noindent {Finally, we re-analyzed the C14 sample using the DR25 providing a homogeneous catalog with the global seismic properties of the main-sequence and sub-giant stars with detection of solar-like oscillations observed in short-cadence during the survey phase of the {\it Kepler} mission (see Table~\ref{tab2}). To that sample we also added a few tens of stars from \citet{2011A&A...534A...6C,2011ApJ...733...95M,2012A&A...543A..54A,2015A&A...582A..25A,2017A&A...601A..82W}. In 19 stars we have realized that the SAP aperture of the new DR25 is not appropriate for asteroseismic studies because they were too small (2 or 3 pixels). By applying the same aperture extraction methodology described in this paper for the LC dataset we were able to reduce the noise level of the resultant light curve and the modes were detectable. In total, we detected the modes for 526 stars with 514 stars in common with the C14 sample. As shown by \citet{2017EPJWC.16001007S} who analyzed a subsample of solar analogs with DR23 and DR25, for the high SNR targets, the global seismic parameters remain the same. The values reported by C14 and the A2Z results with the DR25 have a median offset of 0.5\% in $\nu_{\rm max}$ and of 0.2\% in $\Delta \nu$. {We note that we also report $\nu_{\rm max}$ values for 34 stars that were in C14 but for which only $\Delta \nu$ was provided and no $\nu_{\rm max}$ was given.} We do not report the seismic parameters for 4 stars from C14 as we could not detect solar-like oscillations. }




\section{Granulation and modes amplitudes}


\noindent The determination of the global seismic parameters provides invaluable information on the convective parameters of the solar-like stars as well as the maximum amplitude of the modes. It has been shown that there are tight relations between these parameters and the frequency of maximum power of the acoustic modes \citep[e.g.][]{2011ApJ...743..143H,2011ApJ...741..119M,2012A&A...537A..30M,2014A&A...570A..41K}. Here we study these parameters and compare them with the known sample of solar-like stars with detected oscillations. This comparison allows us to check any deviation from those known relations. 

\subsection{Convective background parameters}

\noindent The study of a large number of stars with solar-like oscillations from the main sequence to the red-giant branch showed that the convective background parameters are correlated with the location of the modes and follow scaling relations \citep{2011A&A...529L...8K,2011ApJ...741..119M,2014A&A...570A..41K}. 


\noindent {\color {black} The background fit was performed with a Monte Carlo Markov Chains (MCMC) strategy using the \texttt{apollinaire} module\footnote{The source code is available at \url{https://gitlab.com/sybreton/apollinaire}} (Breton et al. in prep) which uses the \texttt{emcee} package \citep{2013PASP..125..306F}. 
For the fit, we considered the PSD above 5 $\mu$Hz. The model is composed of four components: a power law for the magnetic activity, two Harvey models \citep{1985ESASP.235..199H} for different scales of granulation, and the photon noise at high frequency. All the results have been visually inspected and some PSD were fitted again with a frequency cut at 50 $\mu$Hz. For that higher frequency cut, the fitted model was only constituted of two Harvey models and a flat noise profile. } {  The Harvey law fitted has the form:

\begin{equation}
H(\nu) = \frac{P_{\rm gran}}{1+(\nu/\nu_{\rm gran})^\alpha},
\end{equation}

\noindent {  where $P_{\rm gran}$ is the granulation power, $\nu_{\rm gran}$ is the characteristic frequency}, and the slope $\alpha$ is fixed to 4 following \citet{2014A&A...570A..41K}. {  $P_{\rm gran}$ is related to the granulation amplitude as the $A_{\rm gran}^2$ per frequency bin.}
}


\noindent Figure~\ref{taugran_numax} shows the granulation frequency for the second Harvey model of convection (the one below the modes), {  $\nu_{\rm gran}$  (top panel)}, and the granulation {  power, $P_{\rm gran}$} (bottom panel), as a function of $\nu_{\rm max}$. The grey points represent the values for a subsample of 163 stars from the C14 sample. The analysis of that subsample was also done with the \texttt{apollinaire} pipeline on data calibrated with the KADACS software that generated the KEPSEISMIC lightcurves. This allows us to directly compare the parameters of the {  stars in our} sample with {  those} of  C14.
We can see in both plots that the new confirmed detections have the convective parameters in the same range as the known oscillating solar-like stars and have similar trends following the relations derived for stars from the main sequence to the red-giant branch \citep{2014A&A...570A..41K} shown with blue dashed lines. For $P_{\rm gran}$, we used the power law relation with a slope of 2.1 {  as given in Section 5.4 of \citet{2014A&A...570A..41K}.} 

\begin{figure}[htbp]
\begin{center}
\includegraphics[width=9cm]{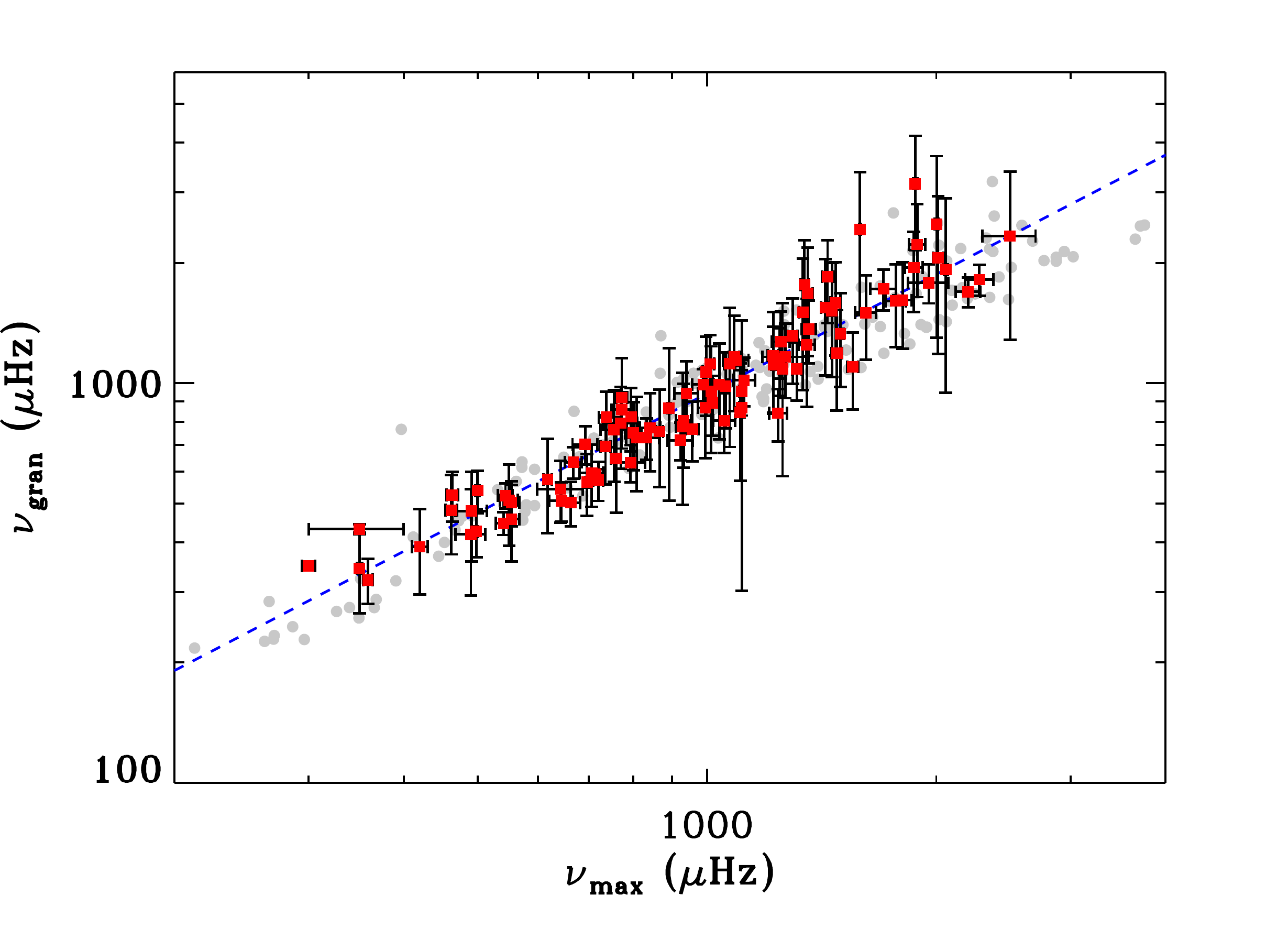}\\
\includegraphics[width=9cm]{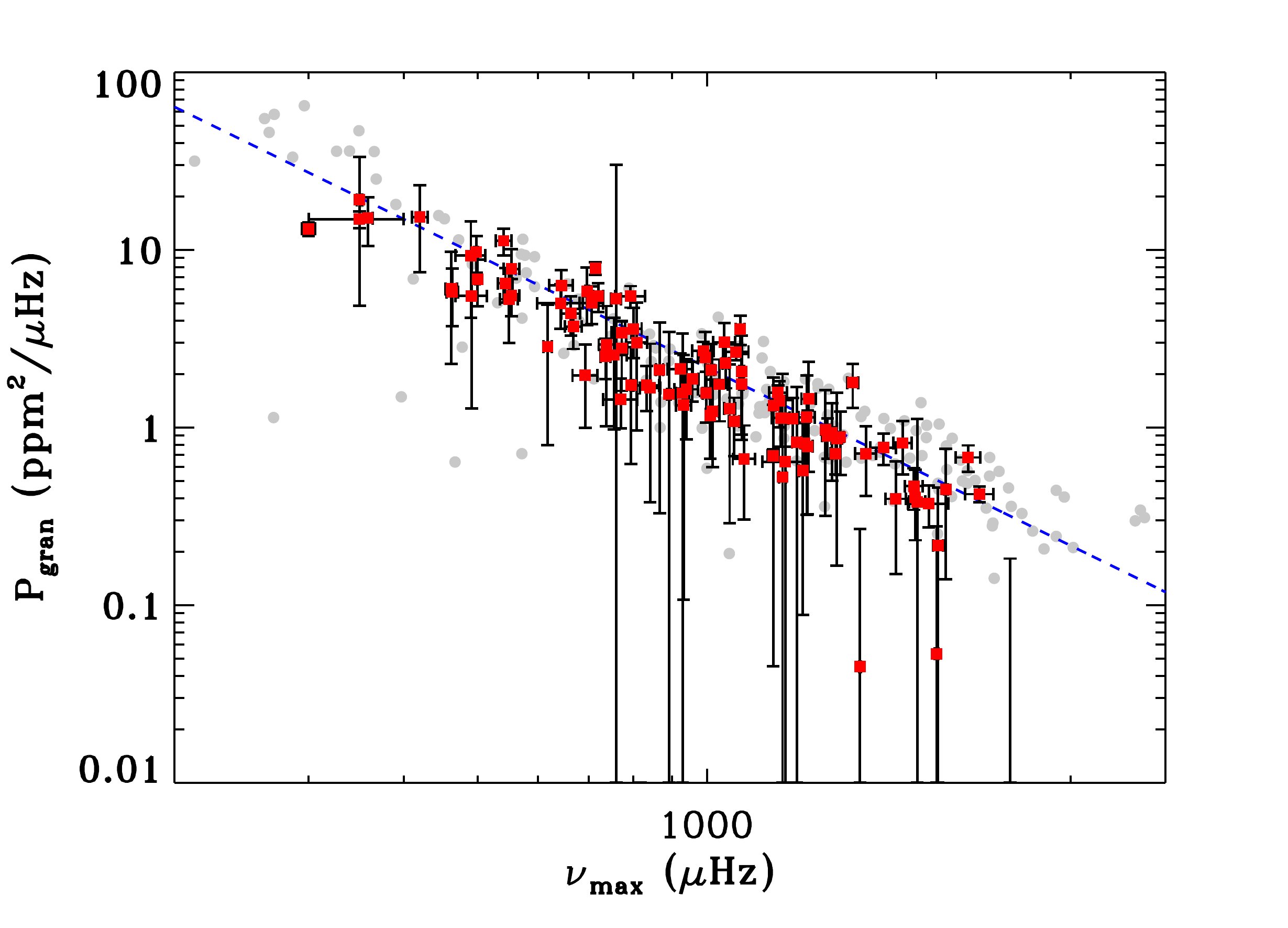}
\caption{Granulation frequency (top panel) and {  power} (bottom panel) as a function of the frequency of maximum power. The {  stars in our sample} are represented with red squares while the subsample of 163 stars from C14 are represented with grey circles. {  The blue dashed lines are the relations derived by \citet{2014A&A...570A..41K}.}}
\label{taugran_numax}
\end{center}
\end{figure}


\subsection{Maximum amplitude of the modes}


From the Gaussian fit around the region of the modes, we derived the bolometric maximum amplitude of the modes following conversion of \citet{2008ApJ...683L.175K} and \citet{2011A&A...531A.124B}. For three stars, namely KIC~3633538, 9529969, 10340511, the Gaussian fit could not converge due to the low SNR and did not provide the maximum amplitude of the modes. {   However, a value for the frequency of maximum power was obtained from the maximum around the region of the modes in the power spectrum density that was smoothed with a boxcar average of width 2\,$\times \Delta\nu$.}

\noindent In Figure~\ref{Amax_numax}, we can see here again that the {  stars in our sample} have a similar behavior compared to the C14 sample. {  We note that for the most evolved stars of our sample -- evolved subgiants with $\nu_{\rm max}$ down to 500\,$\mu$Hz and early red giants with lower $\nu_{\rm max}$  \citep[see Fig.~1 of][]{2014A&A...572L...5M}--, $A_{\rm max}$ is closer to the lower edge. The blue dashed line corresponds to the fit of the form of a power law between $A_{\rm max}$ and $\nu_{\rm max}$ for the C14 sample. We find $A_{\rm max} \propto \nu_{\rm max}^{-0.95}$.} These lower amplitudes combined {  with} the noisier DR24 data, could explain the difficulty to detect the modes in the early analysis of those targets by \citet{2011Sci...332..213C}. 

\begin{figure}[htbp]
\begin{center}
\includegraphics[width=9cm]{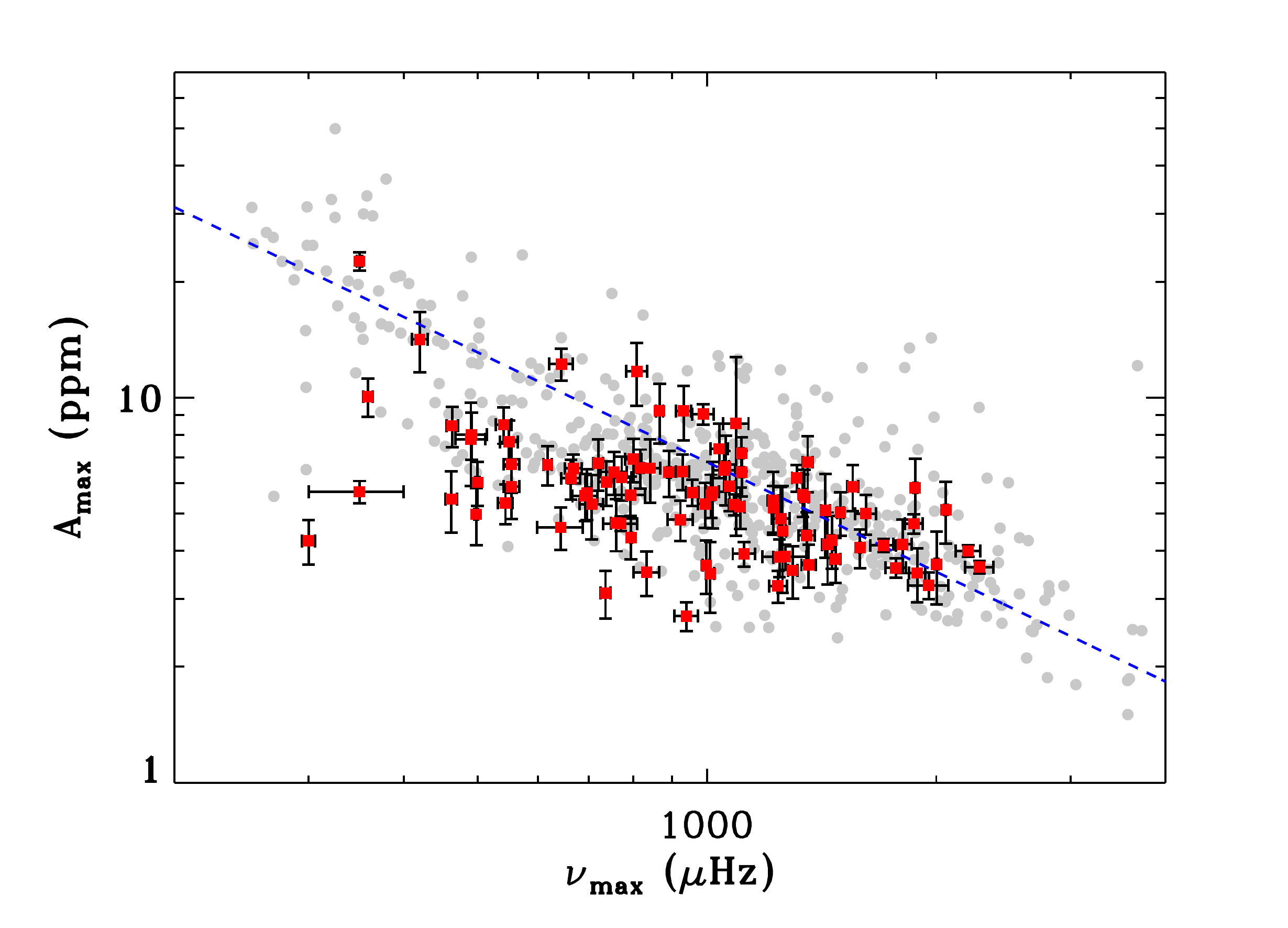}
\caption{Bolometric maximum amplitude of the modes as a function of $\nu_{\rm max}$. The legend is the same as in Figure~\ref{taugran_numax}. {  The blue dashed line corresponds to the fit for the stars from C14.}} 
\label{Amax_numax}
\end{center}
\end{figure}





\section{Global stellar parameters}\label{sec:MR}


\noindent Using the seismic scaling relations \citep{1991ApJ...371..396B,2011A&A...529L...8K}, we computed the masses and radii of {  our sample} where we combined the global seismic parameters and the atmospheric parameters from Section~\ref{sec:atmosph} {  as follows:} 

\begin{equation}
\frac{R}{R_\odot}= \left( \frac{\Delta\nu_{\odot}}{ \Delta \nu}\right)^{-2} \left(\frac{\nu_{\rm max}}{\nu_{\rm max, \odot}} \right) \left(\frac{T_{\rm eff}}{5777} \right)^{1/2}\\
\end{equation}

\begin{equation}
\frac{M}{M_\odot}=\left(\frac{\Delta\nu_{\odot}}{\Delta \nu} \right)^{-4} \left(\frac{\nu_{\rm max}}{\nu_{\rm max, \odot}} \right)^3 \left(\frac{T_{\rm eff}}{5777} \right)^{3/2}. \\
\end{equation}

\noindent {  where $\nu_{\rm max, \odot}$\,=\,3100\,$\mu$Hz and $\Delta\nu_{\odot}$\,=\,135.2\,$\mu$Hz.}

\noindent Among the confirmed detections, {\color {black} 79 stars} have APOGEE spectroscopic parameters and {\color {black} 16} have LAMOST values. The remaining {\color {black} 4} stars have $T_{\rm eff}$ and [Fe/H] from the {\it Gaia-Kepler} catalog.


\noindent In Figure~\ref{MR}, we represent the mass-radius diagram with {  our sample} with red squares and the previously known sample of solar-like stars with oscillation detections in \kep\ with grey circles. For the latter sample, masses and radii were computed by \citet[][hereafter S17]{2017ApJS..233...23S} who used grid-based modeling where the global seismic parameters ($\Delta \nu$ and $\nu_{\rm max}$) were combined {  with} spectroscopic observables ($T_{\rm eff}$, $\log$\,g, and [Fe/H]). We can see that our new sample includes a larger fraction of massive stars as suggested above. 



\noindent Given that the observed $\Delta \nu$ might not be obtained in the asymptotic regime as it should be measured in a frequency range above $\nu_{\rm max}$, several approaches have been developed to apply a correction on the observed mean large frequency separation in order to reduce the impact on the derivation of the stellar masses and radii {  \citep[e.g.][]{2011ApJ...743..161W,2013A&A...550A.126M,2016MNRAS.460.4277G,2016ApJ...822...15S,2018A&A...616A.104K,2021A&A...648A.113B}}. For instance, \citet{2013A&A...550A.126M} derived an empirical correction by taking into account the second-order term in the Tassoul relation \citep{1980ApJS...43..469T} and by using seismic observations for main-sequence to red-giant stars. The approach by \citet{2016ApJ...822...15S} used a grid of models of red giants where the mean large frequency from the the sound-speed profile was compared to the one obtained by using the computed frequencies, leading to a grid of $\Delta \nu$ corrections for each model of the grid. Given that our targets have a rather small SNR and that they are main-sequence stars and subgiants, we decided to adopt the \citet{2013A&A...550A.126M} corrections. They led to smaller radii by 1.5\% {  on} average and smaller masses by 2.9\% {  on} average compared to the radii and masses obtained without $\Delta \nu$ corrections.

\begin{figure}[htbp]
\begin{center}
\includegraphics[width=9cm]{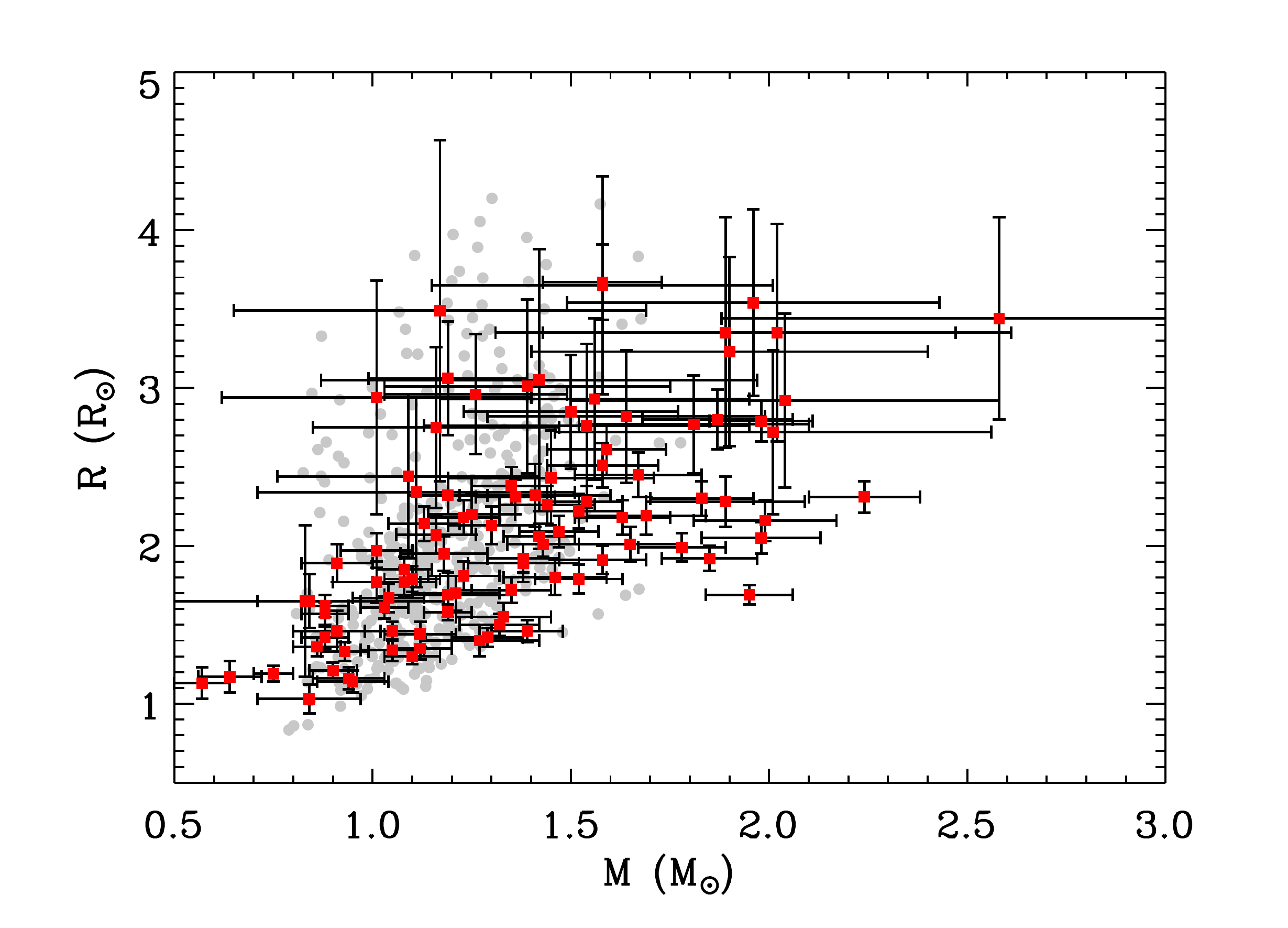}
\caption{Radius vs Mass diagram for the C14 stars using S17 parameters (grey circles) and the {  stars in our sample} described in this work (red diamonds). For the {  latter}, corrections on $\Delta \nu$ were applied following Mosser et al. (2013).}
\label{MR}
\end{center}
\end{figure}

\noindent In Figure~\ref{MR}, we can see that 2 stars have masses above 2\,$M_\odot$, which are not expected in our sample {  as they are in the instability strip and do not have an external convective envelope needed for the excitation of solar-like oscillations \citep{2010aste.book.....A}}. Given the large uncertainties on the seismic parameters that are reflected by the larger error bars for those stars they could be attributed to the low SNR of the modes. 

\section{Rotation and magnetic activity}\label{sec:rot}

For the {\color {black} 99} stars with a confirmed detection of solar-like oscillations, we analyzed the long-cadence time series to look for the surface rotation periods and measure the level of magnetic activity. These measurements are based on the presence of active regions/spots that come in and out of view and lead to a modulation in the lightcurves that is related to the surface rotation. For this analysis we applied three different techniques: a time-frequency analysis based on wavelets \citep{1998BAMS...79...61T, liu2007,2010A&A...511A..46M}, the auto-correlation function \citep{2014A&A...572A..34G,2014ApJS..211...24M}, and the composite spectrum that combines the two previous methods \citep{2016MNRAS.456..119C,2017A&A...605A.111C,2019ApJS..244...21S}. The reliable rotation periods, $P_{\rm rot}$, were selected following criteria described in \citet{2019ApJS..244...21S}, \citet{2021A&A...647A.125B} and \citet{2021ApJS..255...17S}. {  Briefly, we use the three different filtered KEPSEISMIC lightcurves to select the rotation period. We then use the PDC-MAP light curves to confirm that the rotational signal does not result from pollution (instrumental or stellar) as the KEPSEISMIC lightcurves have a larger aperture compared to the PDC-MAP and are calibrated without taking into account the information on the instrumental drifts embedded in the co-trending basic vectors used in PDC-MAP. We should note that PDC-MAP light curves are often filtered at 20 days and, in \citet{2019ApJS..244...21S}, we showed that solely by using PDC-MAP light curves the distribution of retrieved rotation periods is shifted towards shorter values. When the rotation periods inferred from both calibration systems agree inside the combined errors, {  they are} less likely to come from some instrumental pollution {  or pollution by nearby stars}. Some examples of lightcurves are shown in Appendix D.} We obtained surface rotation periods for {\color {black} 63} stars among which {\color {black} one} was classified as a close binary candidate {  based on the fast rotation and the shape of the lightcurve \citep[See][for more details]{2019ApJS..244...21S,2021ApJS..255...17S}}. 

\noindent For stars with rotation period measurements, we computed the photometric magnetic index, $S_{\rm ph}$, following \citet{2014JSWSC...4A..15M}. Briefly, we first calculate the standard deviation of  subseries of length 5\,$\times P_{\rm rot}$ and compute the mean value. This index has been shown to provide a relevant proxy for magnetic activity based on the Sun and a few solar-like stars and comparison with classical magnetic activity indexes \citep{2016A&A...596A..31S,2017A&A...608A..87S}.

\begin{figure}[htbp]
\begin{center}
\includegraphics[width=9cm]{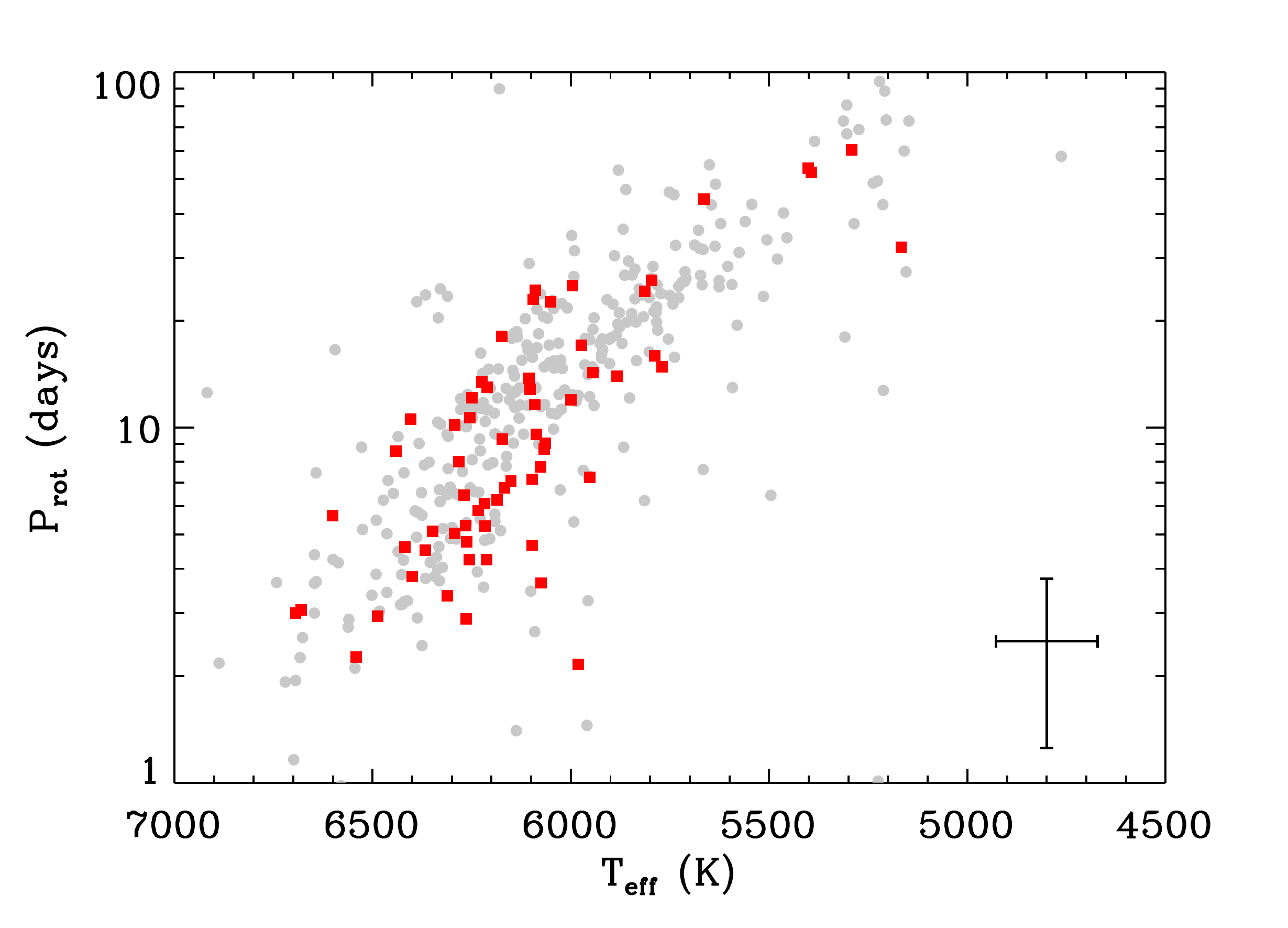}
\caption{Surface rotation periods, $P_{\rm rot}$, as a function of effective temperature, $T_{\rm eff}$, for the G14 sample (grey circles) and {  our sample} (red squares). Typical error bars are shown in the bottom right corner.}
\label{Teff_Prot}
\end{center}
\end{figure}

\noindent In Figure~\ref{Teff_Prot} we represent $P_{\rm rot}$ as a function of the effective temperature of the stars for the new seismic detections and the sample of   \citet[][hereafter G14]{2014A&A...572A..34G}. The stars with new seismic detection seem to be quite homogeneously distributed. While we had noticed that a significant fraction of those stars are in their subgiant phase, very few stars have long rotation periods. Because stars spin-down as they evolve {  \citep[e.g.][]{1972ApJ...171..565S,1988ApJ...333..236K,2013A&A...556A..36G}}, a larger number of slow-rotating  ($P_\text{rot}>40$ days) subgiants were expected in comparison with what we found in this work. However many of the subgiants of our sample are in general more massive than 1.2\,M$_\odot$, which corresponds to the Kraft break \citep{1967ApJ...150..551K}. Stars above the Kraft break do not undergo magnetic braking due to stellar winds in contrast with the lower mass stars. This difference between the high-mass and low-mass solar-like stars could explain the rotation periods retrieved for our sample.




\section{Discussion}




We look here at the different properties of the stars with {  detections} of solar-like oscillations {  presented in this paper} and compare them to the C14 sample, in particular to investigate a possible gain in the parameter space that we are probing. 

\subsection{Magnitude distribution}

We first look at the magnitude of the targets. Indeed, in addition to the calibration issue of the data, the previous non detection of the modes for those stars could also be affected by higher photon noise that depends on the stellar magnitude. The higher noise in the data could lead to a non detection. Figure~\ref{Kp_histo} shows the distribution of the \kep\, magnitude, $K_p$, for {  our sample} and the C14 sample. While they probe a similar range of magnitudes, we can see that the {  stars in our sample} have a larger fraction of stars fainter than 11 compared to the C14 sample. We find that {\color {black} 55.5\%} of the stars in our sample have $K_p\,>$\,11, compared to 39\% for the C14 sample.

\noindent We perform a Kolmogorov-Smirnov test to quantify the comparison of the distribution of the magnitudes between the C14 sample and our sample of new seismic detections. We find a deviation value of {\color {black} 0.22} as well as a probability on the distributions differences of {\color {black} 0.05\%}, which means that the magnitudes distributions are indeed very different. This confirms that our sample with {  detections} of solar-like oscillations {  presented in this paper} are fainter.

\begin{figure}[htbp]
\begin{center}
\includegraphics[width=9cm]{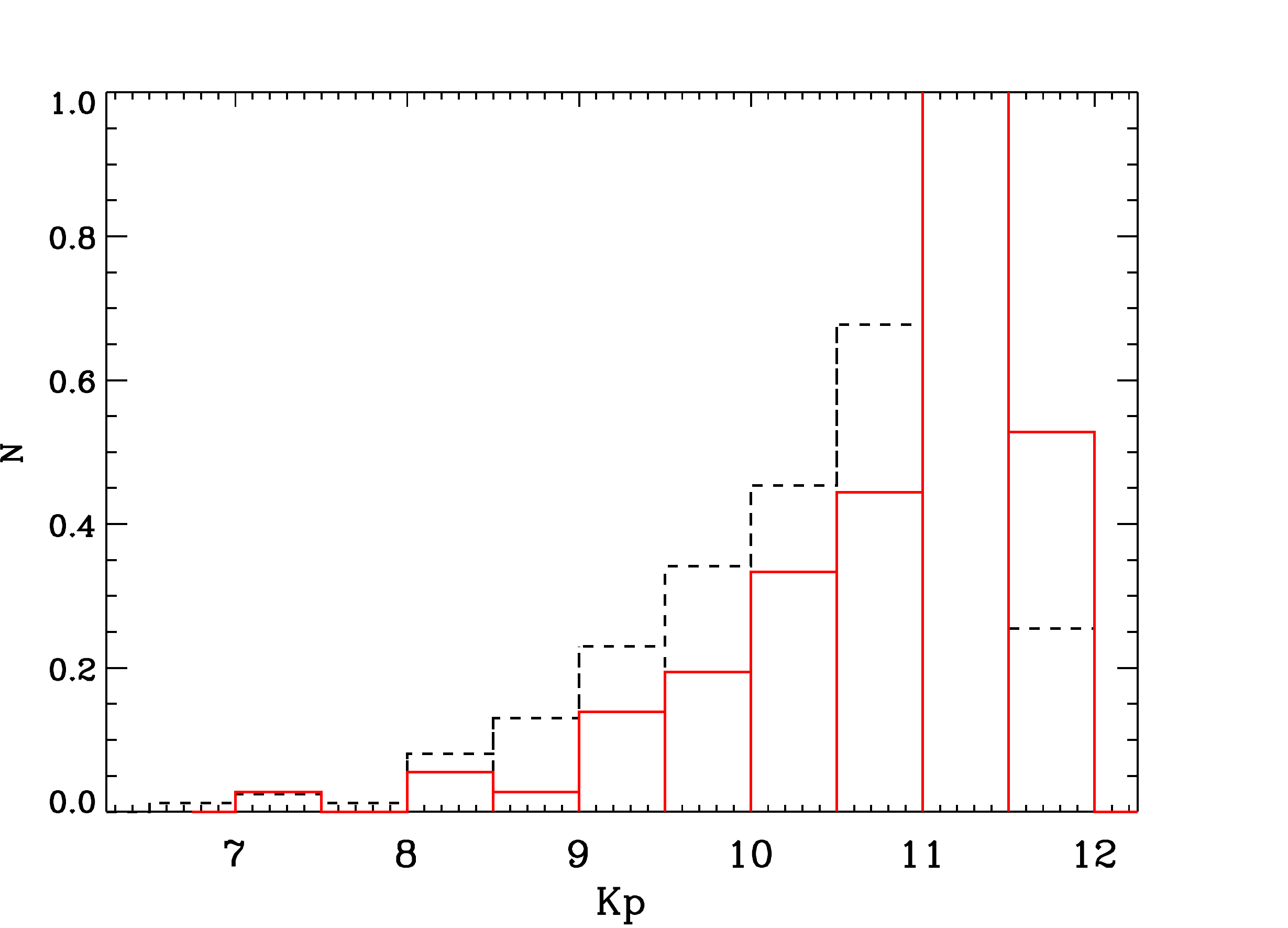}
\caption{Normalized histogram of the \kep\, magnitude for the C14 sample (black {  dashed} line) and {  our sample} (red solid line). }
\label{Kp_histo}
\end{center}
\end{figure}

\subsection{Blends explaining the previous non detections}
With these {  detections} of solar-like oscillations {  in our sample} using the new DR25, we can wonder whether the previous non detection was related to the presence of nearby stars. To check for that possibility, we cross-matched the list of {  stars in our sample} with Gaia EDR3 \citep{2021A&A...649A...1G} and checked two possible sources of amplitude dilution: 
\begin{enumerate}
\item Companions that are unresolved by {\it Gaia}, as indicated by high {\it Gaia} RUWE (Renormalized Unit Weight Error) values (RUWE$>$1.4 typically indicates binaries). While the contrast  between the main star and the companions is not known, it is possible that a pollution can arise.
\item All resolved stars in {\it Gaia} within 20\arcsec (5 {\it Kepler} pixels) of the target. For stars with multiple companions, we selected the brightest star within the search radius.
\end{enumerate}

\noindent From that analysis, {  we find 15 stars where RUWE\,$>$\,1.4 and 9 stars for which the magnitude difference between our target and the brightest star within 20\arcsec\, is smaller than 3}. So 24 stars are flagged to have a nearby star that may dilute the amplitude of the flux and could have prevented the detection of the modes. This number represents an upper limit of the number of stars for which oscillations could not be detected due to a blend. We also checked the amplitude of the modes for those stars that we flagged and found that 7 of them have a lower amplitude of the modes, which can thus be explained by the possible blends. 
 
\noindent {  For the remaining 22 stars without a nearby polluting star it is very likely that the better quality of the data allows us to detect the modes presented in this paper.}

\subsection{Metallicity distribution}

Metallicity can play a role on the detectability of solar-like oscillations as was shown by \citet{2010A&A...509A..15S}. Indeed, they suggested that for metal-poor stars, the amplitude of the modes was decreased. Figure~\ref{FeH_histo} shows the distribution of the metallicity for the {\color {black} 95} stars {  of our sample} (red solid line), where [Fe/H] comes from either APOGEE or LAMOST as metallicity from B20 is not as reliable. We can see that it peaks at {\color {black} solar metallicity}. For comparison the metallicity distribution of the C14 sample is represented on the same figure with the {  dashed} line, also peaking at solar metallicity. We note that among the {  stars in our} sample, {\color {black} 51\%} of {  them} are metal poor compared to the Sun while 54\% of the C14 sample are metal poor but given the typical uncertainties of 0.1dex the difference does not seem significant. We quantify the differences between the distributions of the two sample using a Kolmogorov-Smirnov test. We obtain a deviation of {\color {black} 0.08} with a probability that one sample is different from the other of {\color {black} 62.4\%}, confirming that the distributions are not significantly different. 

\begin{figure}[htbp]
\begin{center}
\includegraphics[width=9cm]{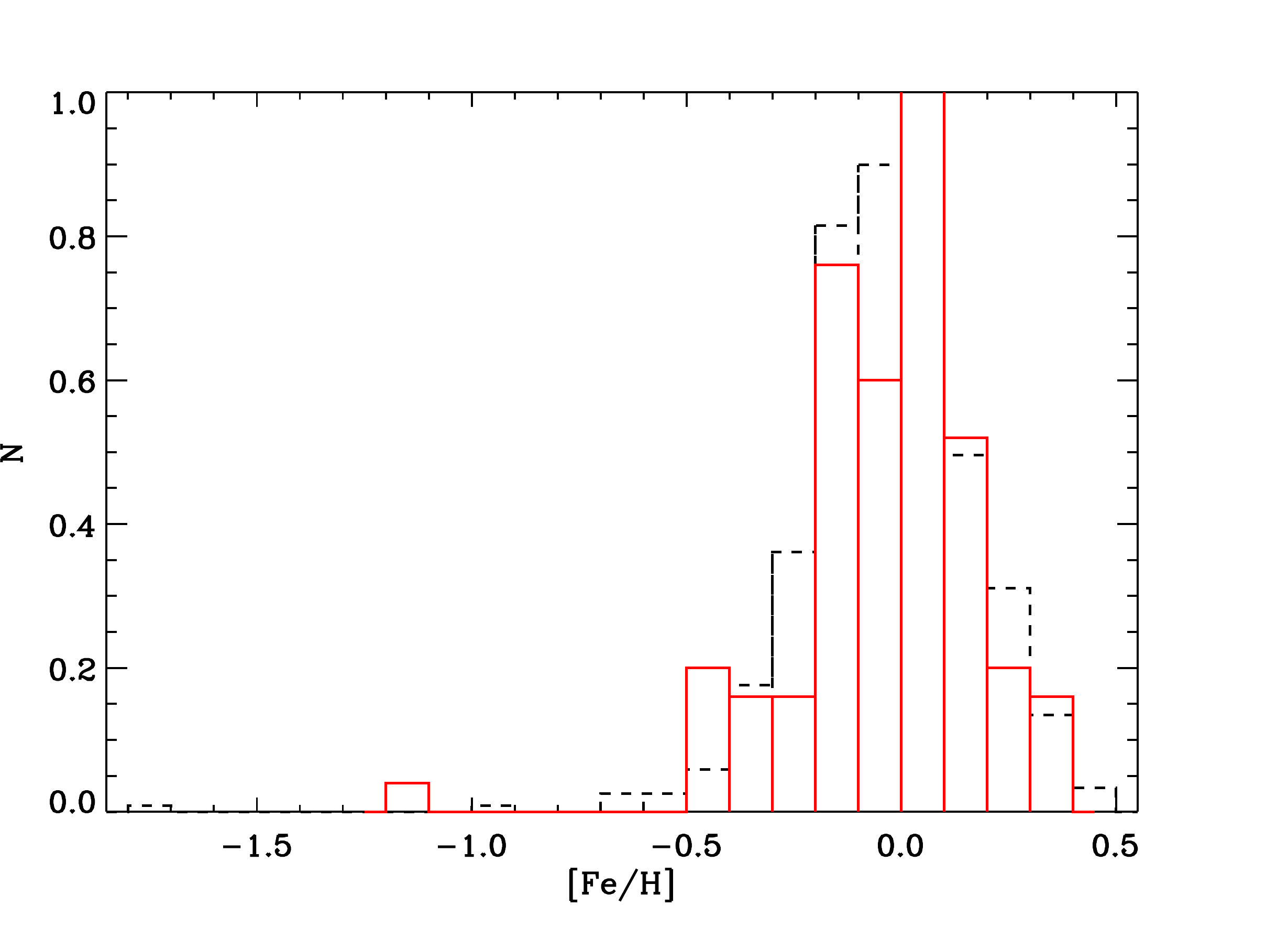}
\caption{Normalized histogram of the \kep\ metallicity for the C14 sample (black {  dashed} line) and {  for our sample} (red solid line) for the 96 stars with new seismic detections and with APOGEE or LAMOST spectroscopic observations. }
\label{FeH_histo}
\end{center}
\end{figure}

\subsection{Surface magnetic activity}

{  As explained in Section~5.2,} from the photometric activity proxy, $S_{\rm ph}$, we can also investigate whether the {  stars with} {  of our sample} have different levels of magnetic activity compared to G14.  In Figure~\ref{Sph_Prot}, we show the magnetic activity index $S_{\rm ph}$ as a function of the rotation period $P_{\rm rot}$ for the stars from G14 with reliable rotation periods and for which asteroseismic detection had been obtained (grey circles). The results for the {  stars in our sample} are shown with the red squares and the close binary {  candidate with a blue square}. We can see that most stars with detected oscillations have similar $S_{\rm ph}$ values that are in the same range as the solar ones between the minimum and maximum of its magnetic cycle {  as computed in \citet{2019FrASS...6...46M} }  (delimited by the dashed lines). There are some stars with magnetic activity levels above the range of the solar cycle for both the previously known sample and {  our sample}. However the new sample of this work adds four stars that are slower rotators ($P_{\rm rot} \ge$ 30 days) and with high $S_{\rm ph}$ values. We also note that one star (KIC~8165738) has a very low $S_{\rm ph}$ value of $\sim$\,10ppm. The modulation seems to be real. We recall that since we do not have the inclination angles of the rotation axis of the stars, $S_{\rm ph}$ values that we measure are a lower limit of the real magnetic activity level. {

\begin{figure}[htbp]
\begin{center}
\includegraphics[width=9cm]{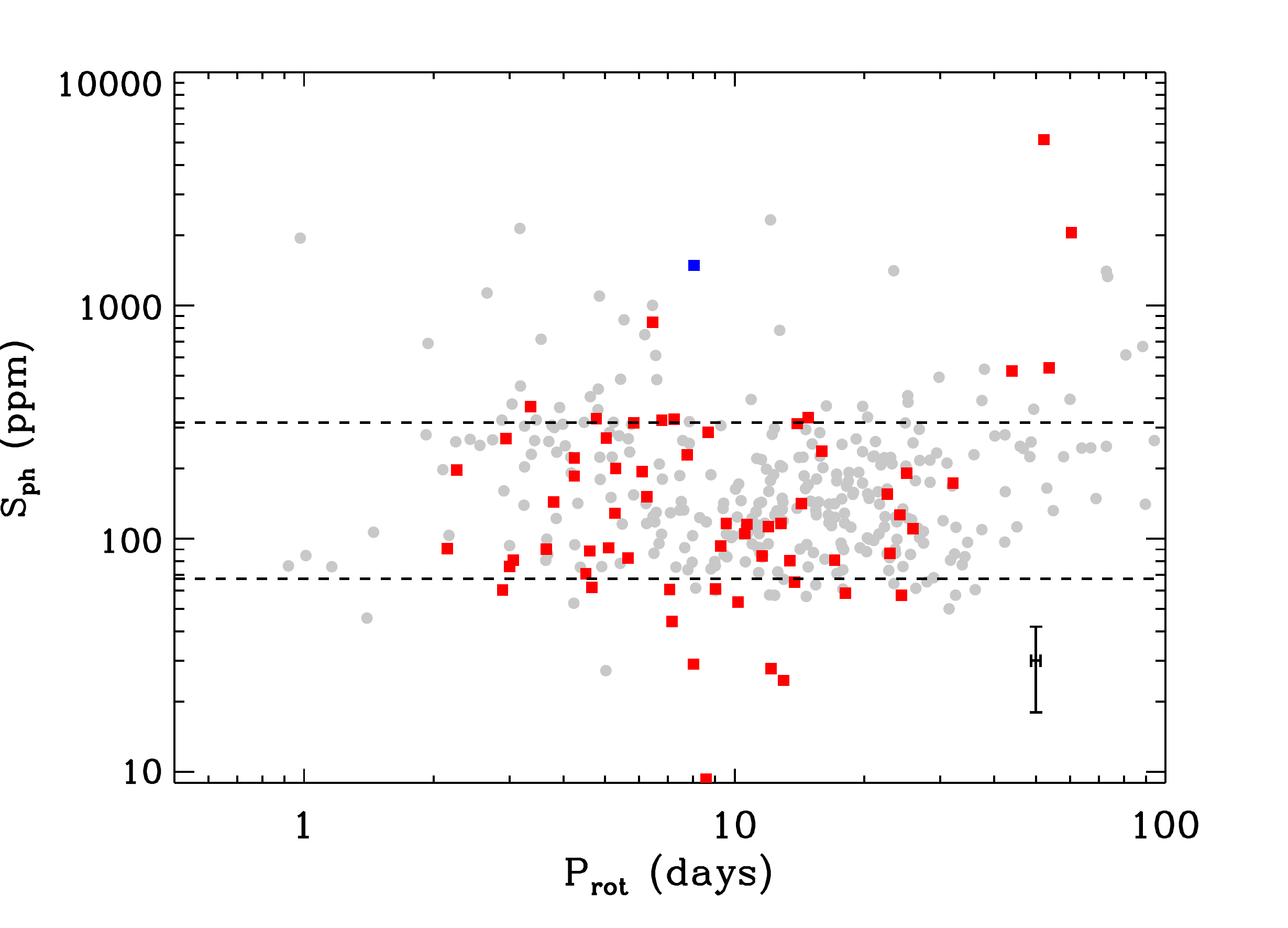}
\caption{Magnetic activity proxy, $S_{\rm ph}$ vs surface rotation period, $P_{\rm rot}$ for the {  G14} sample (grey circles) and {  our sample} (squares). The  blue square is the close binary candidate, KIC 4255487. The dashed lines correspond to the $S_{\rm ph}$ values between minimum and maximum magnetic activity {  from \citet{2019FrASS...6...46M}}, The typical error bars are represented in the bottom right-hand side.}
\label{Sph_Prot}
\end{center}
\end{figure}

\noindent {  Among those stars, three are flagged as potential binary systems as discussed in Section 3.2 (KIC~3633538, KIC~777146, KIC~10969935).} 

\begin{figure*}[h!]
\begin{center}
\includegraphics[width=9cm]{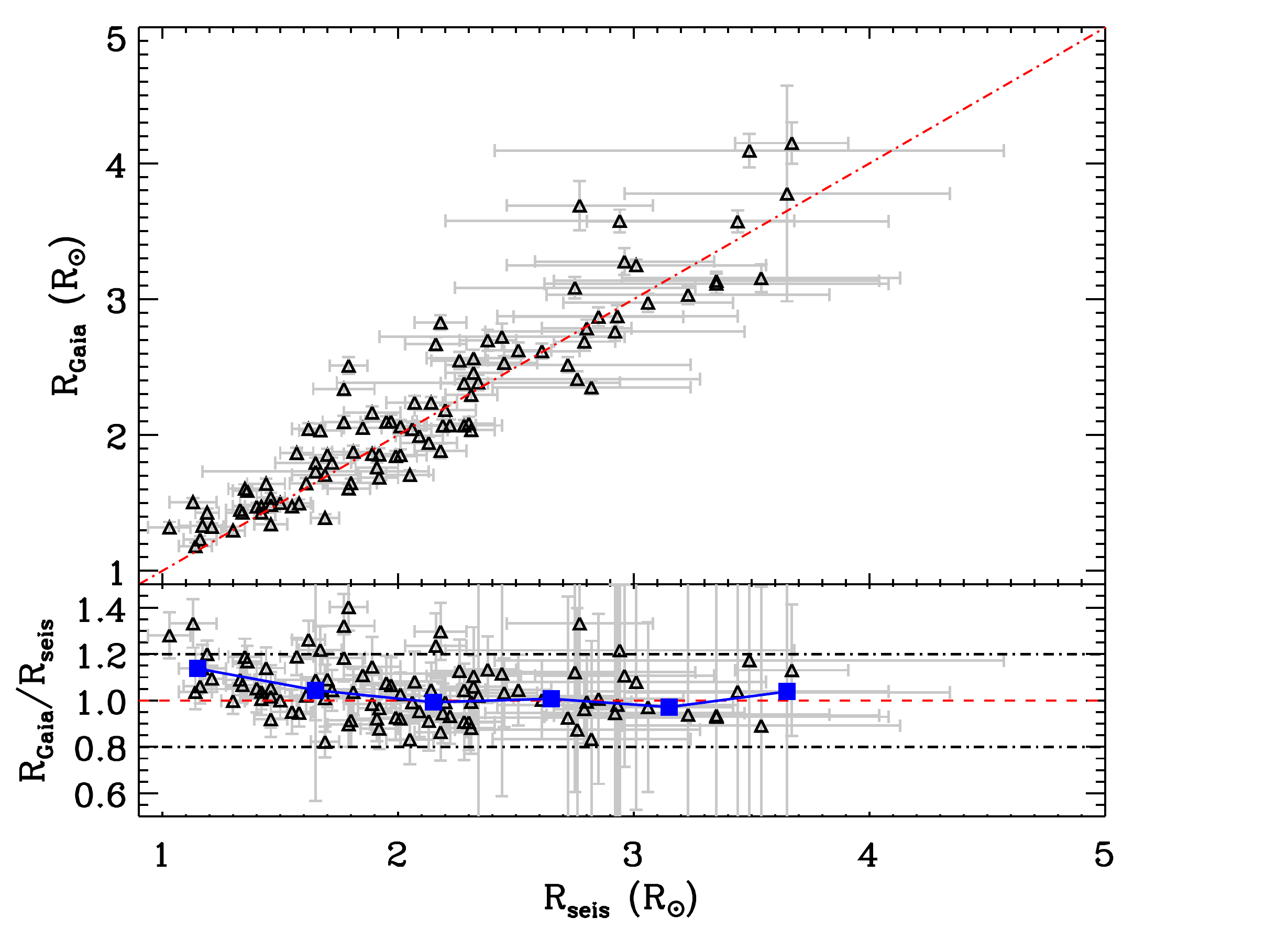}
\includegraphics[width=9cm]{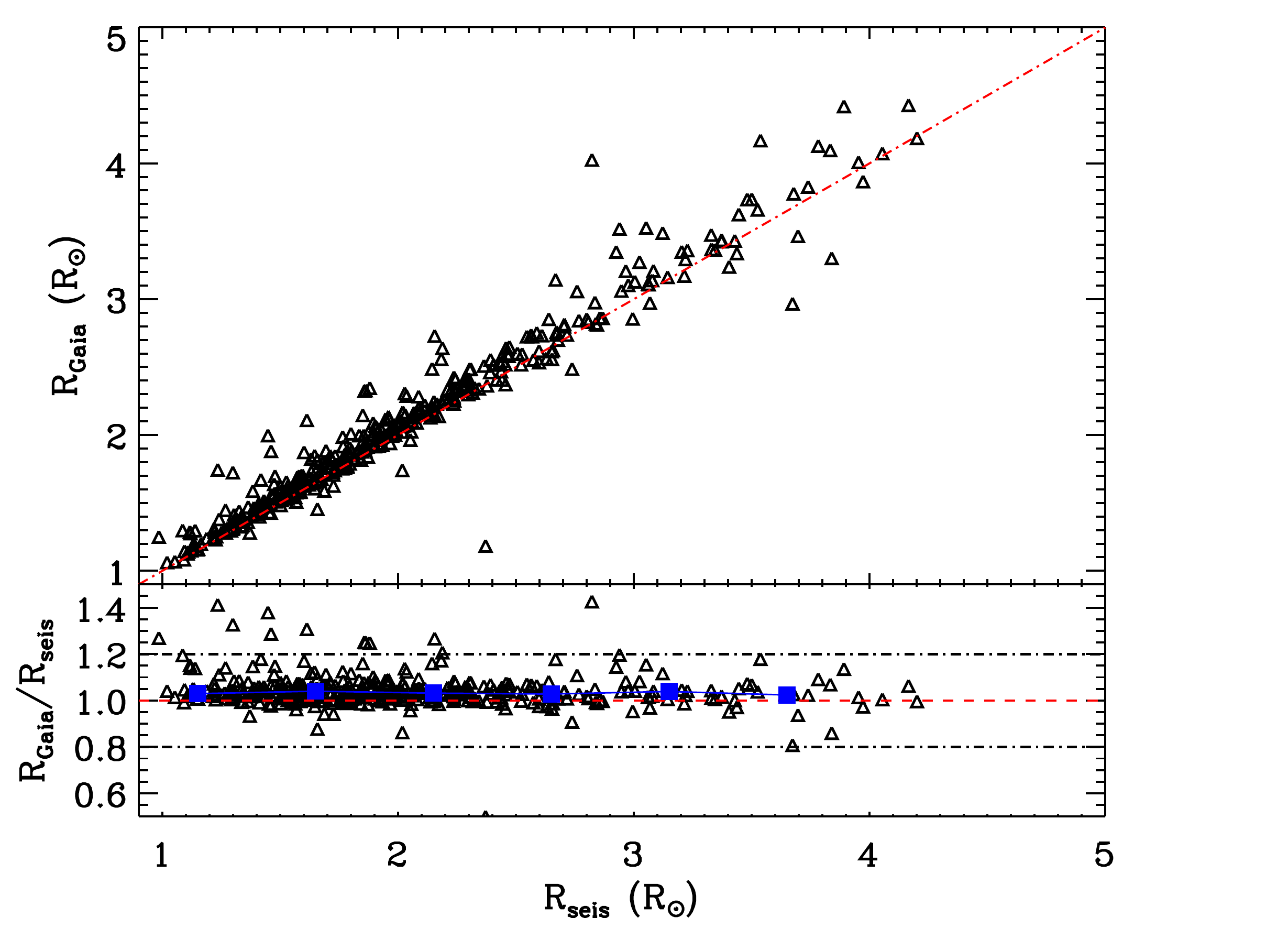}
\caption{Left panel: Comparison between the seismic radii after applying the \citet{2013A&A...550A.126M}  corrections on $\Delta \nu$ and the {\it Gaia} radii (top panel) and ratio of the radii for {  our sample} (bottom panel). {  The blue squares represent the median binned data.} Only 97 stars are shown as two targets do not have a {\it Gaia} radius. Right panel: Same for the S17 sample. }
\label{RGaia}
\end{center}
\end{figure*}

\subsection{Radius comparison with Gaia DR2}


\noindent As shown in Section~\ref{sec:atmosph}, the \kep\,-\emph{Gaia} stellar properties catalog made use of the \emph{Gaia} DR2 in order to improve the stellar parameters of the \kep\ targets by using the new precise parallaxes measurements. B20 fitted isochrones using spectroscopic and photometric information available from previous catalogs and combined them with the \emph{Gaia} data. As said before, 97 of our solar-like stars with new seismic detections are in the B20 catalog.

\noindent We find a general agreement between the seismic radii, $R_{\rm seis}$ and the \emph{Gaia} radii, $R_{Gaia}$, for both  {  our sample} (left panel of Figure~\ref{RGaia}) and the S17 sample (right panel of Figure~\ref{RGaia}) for comparison. For more clarity in the figure, we represent the S17 sample without the uncertainties. In {  our} sample, we note that for larger stars (radii above $\sim$\,2.5$R_{\odot}$), hence more evolved stars,  the seismic radii uncertainties become larger, due to larger uncertainties on the seismic parameters in comparison with smaller stars.  We find that in average the {\it Gaia} radii are overestimated compared to the seismic ones by {\color {black} 4.4\%, with a scatter of 12.3\%. A general trend can be seen, where with increasing seismic radii $R_{Gaia}/R_{\rm seis}$ decreases. For main-sequence stars, the average disagreement is of 10.2\%, where the seismic radii are underestimated compared to the {\it Gaia} radii. For subgiants with radii between 1.5 and 3\,$R_{\odot}$, we find that the seismic radii are underestimated by 3.3\% {  on} average. Finally, for early red giants with radii above 3\,$R_\odot$, the seismic radii are overestimated by 1.2\% {  on} average. }

\noindent These results are slightly different from what was found by \citet{2017ApJ...844..102H} where a similar comparison was done on a larger sample of stars from the main sequence to the red giants but using the Tycho-{\it Gaia} astrometric solution \citep{2015A&A...583A..68M,2016A&A...595A...2G}. Their sample included the previously known solar-like stars with detection of acoustic modes. They found a disagreement of 5\% for radii between 0.8 and 8\,$R_\odot$, where seismic radii for dwarfs and subgiants were underestimated compared to {\it Gaia} radii. Later \citet{2018MNRAS.481L.125S}, \citet{2019A&A...628A..35K}, and \citet{2019ApJ...885..166Z} looked at differences between {\it Gaia} radii obtained with DR2 and the seismic ones, for dwarfs and/or red giants. For red giants,  \citet{2019A&A...628A..35K} reported a 2\% discrepancy between the seismic and astrometric radius and showed how stellar radii can be calibrated with \kep\ and {\it Gaia} (see their Figs. 17 and 19). For dwarfs,  \citet{2018MNRAS.481L.125S} and \citet{2019ApJ...885..166Z} found that the seismic radii were underestimated compared to the {\it Gaia} ones by 2\%. In particular, \citet{2019ApJ...885..166Z} investigated random and systematic uncertainties related to luminosity, effective temperatures, bolometric corrections and estimated the systematic uncertainties to be around 2\% as well. Our findings for the new seismic detections show larger discrepancies. Taking the systematics errors, the results obtained for the new seismic detections are thus consistent with the previously known solar-like stars when using {\it Gaia} DR2.



\noindent To check wether any of the bias or trends between seismic radii and radii inferred with {\it Gaia} observations are due to any effect of the effective temperature scale, we compute the seismic radii using the effective temperatures from B20. We find that {  on} average the {\it Gaia} radii are overestimated by {\color {black} 3.3\%} with a scatter of {\color {black}12.4\%}. Using the effective temperatures from B20 should bring the discrepancy closer to 0. Thus the difference that remains here {  is} likely to be related to the input from the global seismic parameters {  (see Appendix E)}.

\noindent Finally, in order to better assess the agreement between the two ways of obtaining the stellar radii, we compare the radii differences as a function of the combined uncertainties from both seismology and {\it Gaia} (see Figure~\ref{Diff_R_histo}). For {\color {black} 50.5\%} of the stars, the differences between the two methodologies are within 1\,$\sigma$. 

\begin{figure*}[h]
\begin{center}
\includegraphics[width=14cm, height=9cm]{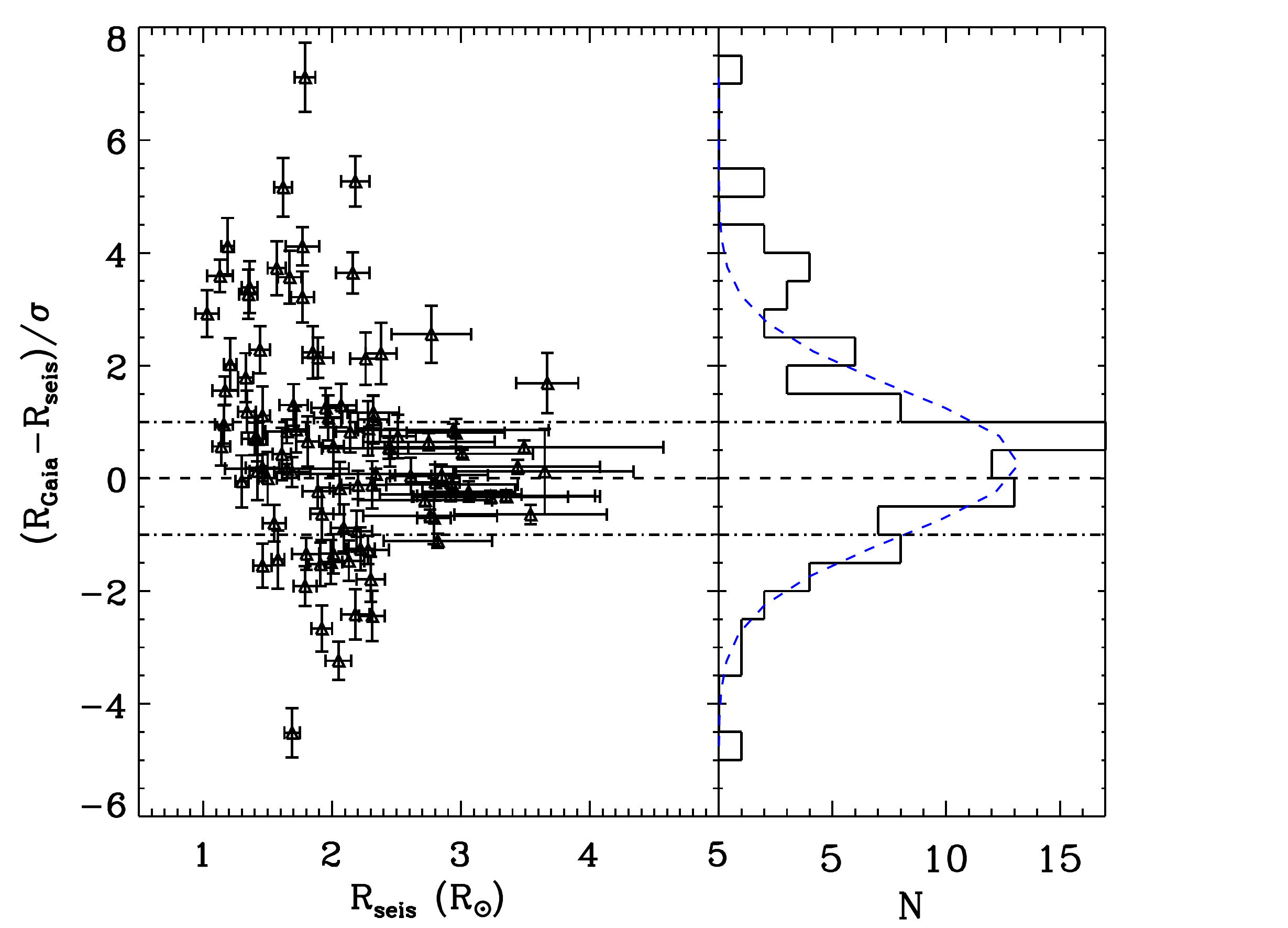}
\caption{Left panel: Differences between the {\it Gaia} radius and the seismic radius in units of the statistical uncertainty ($\sigma$), computed as the quadratic sum of the uncertainties from the B20 {  radii} and the scaling relations {  radii}. Right panel: histogram of the same differences. The {  dashed} lines show the equality of both radii and the dot-dash lines represent the $\pm$\,1\,$\sigma$ limits. Only 97 stars are shown as two targets do not have a {\it Gaia} radius. {  The blue dashed line shows a Gaussian fit of the histogram.}}
\label{Diff_R_histo}
\end{center}
\end{figure*}

\section{Conclusions}
We analyzed the lightcurves from the latest data release of the \kep\ mission for $\sim$\,2,600 stars that were observed in short cadence for one month during the survey phase of the mission. We obtained reliable seismic detections for {  99} solar-like stars, among which {  46} stars with newly reported detections. Our analysis increased by more than 15\% the number of \kep\, dwarfs with a full seismic study. {  We also re-analyzed the DR25 of the sample of stars with previously detected modes by C14 and provide a homogeneous catalog of global seismic parameters for additional 526 main-sequence and sub-giant stars observed by {\it Kepler} during the survey mission, yielding a homogeneous seismic properties catalog for {  625 stars for which a complete table of the global seismic parameters are given in Appendix E}.}

\noindent For the full sample, we consolidated atmospheric parameters from APOGEE, LAMOST, and the \kep\,-{\it Gaia} catalogs. Using the surface gravity and effective temperature from the {\it Kepler-Gaia} catalog, we predicted frequency of maximum oscillation power of the modes based on seismic scaling relations. We found that {  on} average the predicted $\nu_{\rm max}$ is slightly higher than the observed one. 

\noindent The computation of the seismic masses and radii from the scaling relations and with the \citet{2013A&A...550A.126M} corrections on the mean large frequency spacing suggested that {  our target stars on average more massive and off the main sequence compared to the existing \kep\ dwarfs and subgiants in the literature}. 

\noindent While the background parameters have similar trends with $\nu_{\rm max}$ {  as previously found in the literature, the maximum amplitude of the modes appears to be on the lower end of the known trends.} We also found that the sample of new seismic detections is constituted of fainter stars, which can lead to lower amplitude of the modes. This could partly explain the non detection of the acoustic modes for those stars when using the DR24 that was affected by some calibration issues concerning the short-cadence data of \kep. Both our sample and the sample of \cite{2014ApJS..210....1C} have {  on} average a solar metallicity.

\noindent {  By using {\it Gaia} data to look for stars close to our targets, we found that 7 of the stars with {  detections} of solar-like oscillations {  in our sample} have a nearby companion that may pollute the observations. This could explain the non detections of acoustic modes with the previous {\it Kepler} data releases.}

\noindent From the analysis of the surface rotation of the lightcurves, we obtained reliable rotation periods for 64 stars. These stars have magnetic activity levels in the same range as the Sun along its activity cycle. Even though the sample with new seismic detections has a larger number of subgiants we did not find many slow rotators. This is probably because these are massive subgiants that are above the Kraft break on the main sequence and did not go through strong magnetic braking.

\noindent Finally, we found that  seismic radii are {  on} average 4.4\% underestimated compared to the {\it Gaia} DR2 radii with a scatter of 12.3\% and a decreasing trend with evolutionary stage, which is in agreement with previous comparisons done for main-sequence to red-giant stars. Using a different scale of temperature from the {\it Kepler-Gaia} catalog, the discrepancy slightly decreased.

\noindent Many of the stars with new seismic detections have a very low SNR, however around ten of them could have their individual modes characterized, which will be part of a subsequent paper. Even though \kep\ has stopped operating more than five years ago, the high precision and continuous data collected by the mission still represent a goldmine for asteroseismology.

\begin{table*}[h]
\caption{Seismic and stellar parameters of the 99 stars with seismic detections. {  The surface gravity is derived from Eq.\,1.} The Flags provide the provenances of the atmospheric parameters ($T_{\rm eff}$ and [Fe/H]): 0 =APOGEE, 1= LAMOST, 2= {\it Kepler-Gaia} catalog (B20), {  and 3= {\it Gaia} DR2}.}
\begin{center}
\begin{tabular}{ccccccccc}
\hline
\hline
KIC & $T_{\rm eff}$ (K)& $\log g$ (dex)& [Fe/H] (dex)& $\nu_{\rm max}$ ($\mu$Hz)& $\Delta \nu$ ($\mu$Hz)& M (M$_\odot$)& R (R$_\odot$) &  Flag  \\
\hline
2010835 &  5896\,$\pm$\, 126& 4.07\,$\pm$\, 0.01&  0.12\,$\pm$\, 0.15&
 1312\,$\pm$\,   72& 72.73\,$\pm$\,  4.52&  0.91\,$\pm$\,  0.11&
  1.46\,$\pm$\,  0.13&0\\
2578869 &  5395\,$\pm$\, 101& 3.84\,$\pm$\, 0.01&  0.01\,$\pm$\, 0.16&
  809\,$\pm$\,   48& 48.87\,$\pm$\,  1.89&  0.91\,$\pm$\,  0.09&
  1.89\,$\pm$\,  0.12&0\\
3102595 &  5770\,$\pm$\, 122& 3.98\,$\pm$\, 0.01& -0.11\,$\pm$\, 0.02&
 1071\,$\pm$\,   52& 52.20\,$\pm$\,  1.66&  1.83\,$\pm$\,  0.13&
  2.30\,$\pm$\,  0.11&0\\
3124465 &  5796\,$\pm$\, 120& 3.85\,$\pm$\, 0.01& -0.33\,$\pm$\, 0.01&
  795\,$\pm$\,   46& 46.00\,$\pm$\,  1.56&  1.23\,$\pm$\,  0.09&
  2.18\,$\pm$\,  0.11&0\\
3219634 &  6145\,$\pm$\, 133& 3.98\,$\pm$\, 0.02&  0.20\,$\pm$\, 0.01&
 1054\,$\pm$\,   53& 53.86\,$\pm$\,  1.60&  1.69\,$\pm$\,  0.15&
  2.19\,$\pm$\,  0.12&0\\
3238211 &  6092\,$\pm$\, 129& 3.82\,$\pm$\, 0.01& -0.13\,$\pm$\, 0.01&
  720\,$\pm$\,   42& 42.44\,$\pm$\,  1.45&  1.35\,$\pm$\,  0.10&
  2.38\,$\pm$\,  0.12&0\\
3241299 &  6501\,$\pm$\, 130& 4.07\,$\pm$\, 0.01&  0.28\,$\pm$\, 0.01&
 1239\,$\pm$\,   59& 59.70\,$\pm$\,  2.30&  1.99\,$\pm$\,  0.18&
  2.16\,$\pm$\,  0.13&0\\
3425564 &  6239\,$\pm$\, 136& 3.82\,$\pm$\, 0.01& -0.23\,$\pm$\, 0.02&
  714\,$\pm$\,   38& 38.50\,$\pm$\,  5.24&  2.04\,$\pm$\,  0.54&
  2.92\,$\pm$\,  0.55&0\\
3633538 &  5676\,$\pm$\,  85& 4.34\,$\pm$\, 0.03& -0.07\,$\pm$\, 0.01&
 2500\,$\pm$\,  119&119.05\,$\pm$\,  3.91&  0.84\,$\pm$\,  0.13&
  1.03\,$\pm$\,  0.09&3\\
3750375 &  6110\,$\pm$\, 129& 3.69\,$\pm$\, 0.01&  0.04\,$\pm$\, 0.01&
  544\,$\pm$\,   31& 31.16\,$\pm$\,  4.62&  2.02\,$\pm$\,  0.59&
  3.35\,$\pm$\,  0.69&0\\
3761010 &  6403\,$\pm$\, 128& 4.02\,$\pm$\, 0.01&  0.05\,$\pm$\, 0.15&
 1111\,$\pm$\,   59& 59.67\,$\pm$\,  1.89&  1.38\,$\pm$\,  0.09&
  1.92\,$\pm$\,  0.09&1\\
3936658 &  6106\,$\pm$\, 130& 3.70\,$\pm$\, 0.01&  0.10\,$\pm$\, 0.01&
  554\,$\pm$\,   36& 36.66\,$\pm$\,  5.72&  1.09\,$\pm$\,  0.33&
  2.44\,$\pm$\,  0.52&0\\
...\\
\hline

\end{tabular}
\end{center}
\label{tab1}
\end{table*}%

\begin{table*}[h]
\caption{{  Global seismic parameters from A2Z for the {  526} stars with previous detection of solar-like oscillations.}}
\begin{center}
\begin{tabular}{ccc}
\hline
\hline
KIC &  $\nu_{\rm max}$ ($\mu$Hz)& $\Delta \nu$ ($\mu$Hz)  \\
\hline
1430163 &   1807\,$\pm$\,   43& 85.71\,$\pm$\,  1.79\\
1435467 &   1369\,$\pm$\,   56& 70.80\,$\pm$\,  1.47\\
1725815 &   1040\,$\pm$\,   28& 55.97\,$\pm$\,  1.43\\
2010607 &    675\,$\pm$\,    7& 42.44\,$\pm$\,  1.49\\
2309595 &    646\,$\pm$\,   16& 38.97\,$\pm$\,  1.38\\
2450729 &   1078\,$\pm$\,   36& 61.05\,$\pm$\,  1.95\\
2837475 &   1638\,$\pm$\,   72& 75.71\,$\pm$\,  1.22\\
2849125 &    729\,$\pm$\,   31& 40.44\,$\pm$\,  1.29\\
2852862 &    988\,$\pm$\,   61& 54.68\,$\pm$\,  1.07\\
2865774 &   1260\,$\pm$\,   37& 64.20\,$\pm$\,  2.16\\
2991448 &   1127\,$\pm$\,   34& 61.22\,$\pm$\,  2.91\\
2998253 &   2034\,$\pm$\,    9& 89.00\,$\pm$\,  2.13\\
3112152 &   1263\,$\pm$\,   43& 65.00\,$\pm$\,  1.77\\
3112889 &    817\,$\pm$\,   30& 53.09\,$\pm$\,  1.86\\
3115178 &    431\,$\pm$\,   16& 28.72\,$\pm$\,  0.97\\
3123191 &   1704\,$\pm$\,   51& 88.00\,$\pm$\,  2.13\\
3236382 &   1692\,$\pm$\,   15& 73.74\,$\pm$\,  1.80\\
...\\

\hline

\end{tabular}
\end{center}
\label{tab2}
\end{table*}

\begin{table*}[h]
\caption{Rotation periods and magnetic activity levels of the 63 stars with seismic detections. }
\begin{center}
\begin{tabular}{ccc}
\hline
\hline
KIC &  $P_{\rm rot}$ (days) & $S_{\rm ph}$ (ppm)\\
\hline
3633538 &   32.10\,$\pm$\,  4.22&  172.9\,$\pm$\,    4.1\\
6881330 &   11.96\,$\pm$\,  0.70&  112.5\,$\pm$\,    5.6\\
3102595 &   14.80\,$\pm$\,  1.94&  330.2\,$\pm$\,   12.5\\
3124465 &   25.91\,$\pm$\,  1.72&  110.6\,$\pm$\,    3.7\\
3936993 &   10.66\,$\pm$\,  0.85&  115.4\,$\pm$\,    5.8\\
4255487 &    8.04\,$\pm$\,  0.66& 1484.4\,$\pm$\,   67.7\\
4270687 &   13.42\,$\pm$\,  0.81&   80.5\,$\pm$\,    4.8\\
4859338 &    5.83\,$\pm$\,  0.50&  313.6\,$\pm$\,   17.5\\
5112169 &    5.03\,$\pm$\,  0.98&  270.0\,$\pm$\,   17.9\\
5183581 &    6.10\,$\pm$\,  1.23&  193.9\,$\pm$\,   11.1\\
5394680 &   43.94\,$\pm$\,  4.78&  523.7\,$\pm$\,   10.6\\
5597743 &   13.75\,$\pm$\,  1.07&   65.0\,$\pm$\,    3.9\\
5696625 &   10.56\,$\pm$\,  1.33&  105.1\,$\pm$\,    5.9\\
5771915 &    9.27\,$\pm$\,  0.69&   92.9\,$\pm$\,    5.6\\
5791521 &    5.10\,$\pm$\,  0.64&   91.4\,$\pm$\,    6.5\\
5814512 &    5.65\,$\pm$\,  0.67&   82.5\,$\pm$\,    6.5\\
5856836 &    7.06\,$\pm$\,  0.68&   60.5\,$\pm$\,    4.7\\
6062024 &   12.97\,$\pm$\,  2.52&   24.7\,$\pm$\,    2.3\\
...\\

\hline

\end{tabular}
\end{center}
\label{tab1_1}
\end{table*}%

\begin{acknowledgements}
This paper includes data collected by the \emph{Kepler} mission. Funding for the \emph{Kepler} mission is provided by the NASA Science Mission directorate. Some of the data presented in this paper were obtained from the Mikulski Archive for Space Telescopes (MAST). STScI is operated by the Association of Universities for Research in Astronomy, Inc., under NASA contract NAS5-26555. S.~M. acknowledges support by the Spanish Ministry of Science and Innovation with the Ramon y Cajal fellowship number RYC-2015-17697 and the grant number PID2019-107187GB-I00. 
R.~A.~G. and S.~N.~B acknowledge the support from PLATO and GOLF CNES grants. A.~R.~G.~S. acknowledges the support from National Aeronautics and Space Administration under Grant NNX17AF27G and STFC consolidated grant ST/T000252/1. D.H. acknowledges support from the Alfred P. Sloan Foundation, the National Aeronautics and Space Administration (80NSSC19K0597), and the National Science Foundation (AST-1717000). M.S. is supported by the Research Corporation for Science Advancement through Scialog award $\#$26080. Guoshoujing Telescope (the Large Sky Area Multi-Object Fiber Spectroscopic Telescope LAMOST) is a National Major Scientific Project built by the Chinese Academy of Sciences. Funding for the project has been provided by the National Development and Reform Commission. LAMOST is operated and managed by the National Astronomical Observatories, Chinese Academy of Sciences. 
 \end{acknowledgements}
\bibliographystyle{aa} 
\bibliography{/Users/Savita/Documents/BIBLIO_sav}

\newpage

\appendix

\section{Frequency of maximum power comparison with spectroscopic $\log g$}

We computed the predicted frequency of maximum power from the seismic scaling relations (see Eq. 1) using the spectroscopic surface gravities whenever available. Figure~\ref{numax_comp_spec} shows the ratio between the observed and predicted $\nu_{\rm max}$. We can see that the disagreement is larger than when using the $\log g$ from B20, with the predicted values overestimated by 14.6\% {  on} average. 

\begin{figure}[h!]
\begin{center}
\includegraphics[width=9cm]{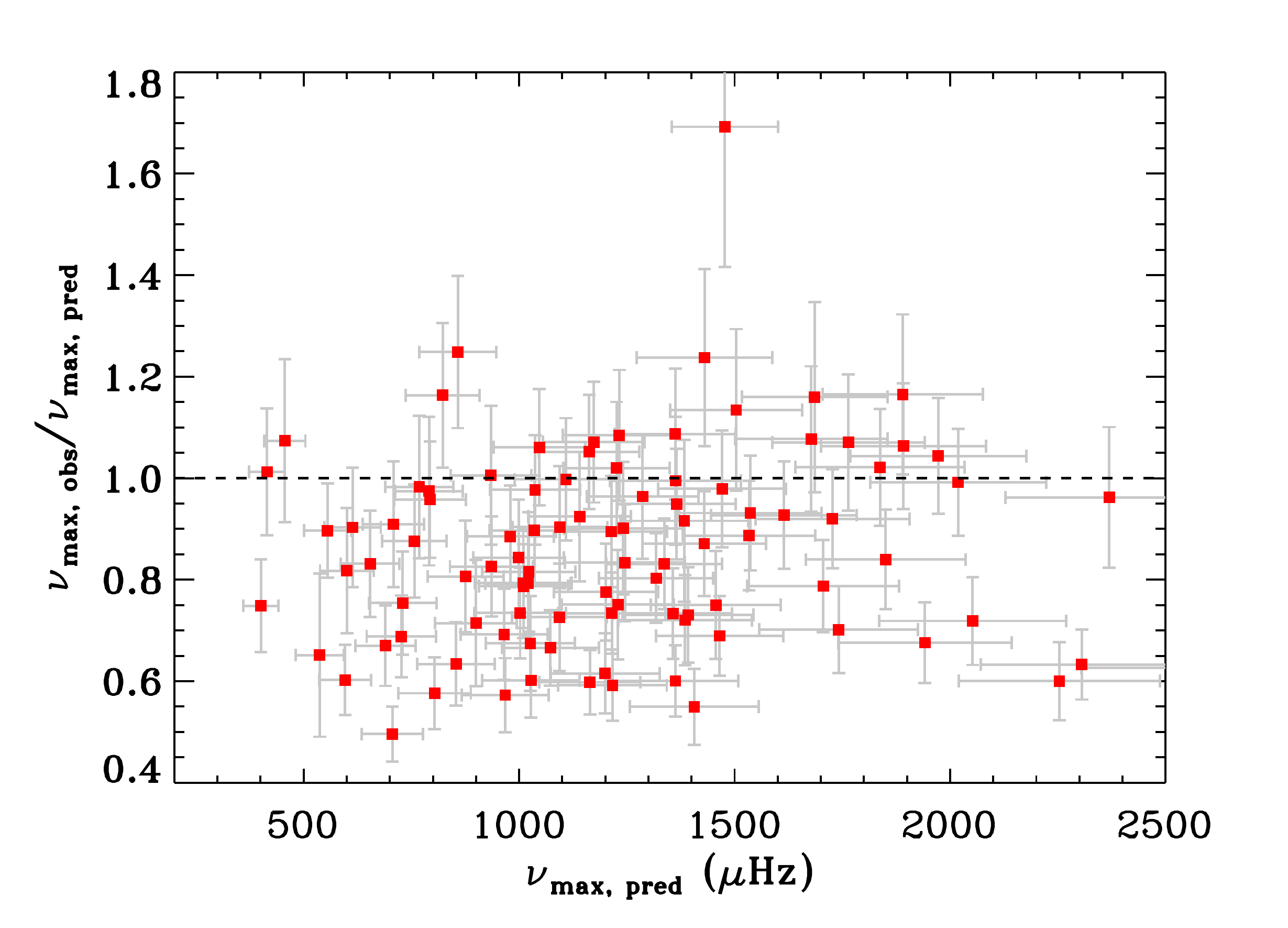}
\caption{Ratio of the observed $\nu_{\rm max, obs}$ and the predicted $\nu_{\rm max, pred}$ using the spectroscopic $\log g$. The {  dashed} line shows the equality of both values.}
\label{numax_comp_spec}
\end{center}
\end{figure}

\section{Candidates for solar-like oscillation detections}

In this appendix, we present the results for the possible detections of solar-like oscillations in 26 targets. As explained in Section~\ref{sec:final}, these targets were not selected because the SNR was too low. In some cases, only one pipeline reported a detection and in other cases the location of the possible $\nu_{\rm max}$ was within 30\% of the predicted value. These candidates do not have a reliable mean large frequency spacing but we report their approximative $\nu_{\rm max}$ in Table~\ref{tab2}. In Figure~\ref{numax_comp_cand}, we show the comparison between the predicted $\nu_{\rm max}$ and the observed one similarly to Figure~\ref{numax_comp}, where we used the surface gravities from B20. Like {  our sample with} confirmed {  detections}, the predicted frequency of maximum power seems to be overestimated but by a larger amount of 12.6\% {  on} average. 

\begin{figure}[htbp]
\begin{center}
\includegraphics[width=9cm]{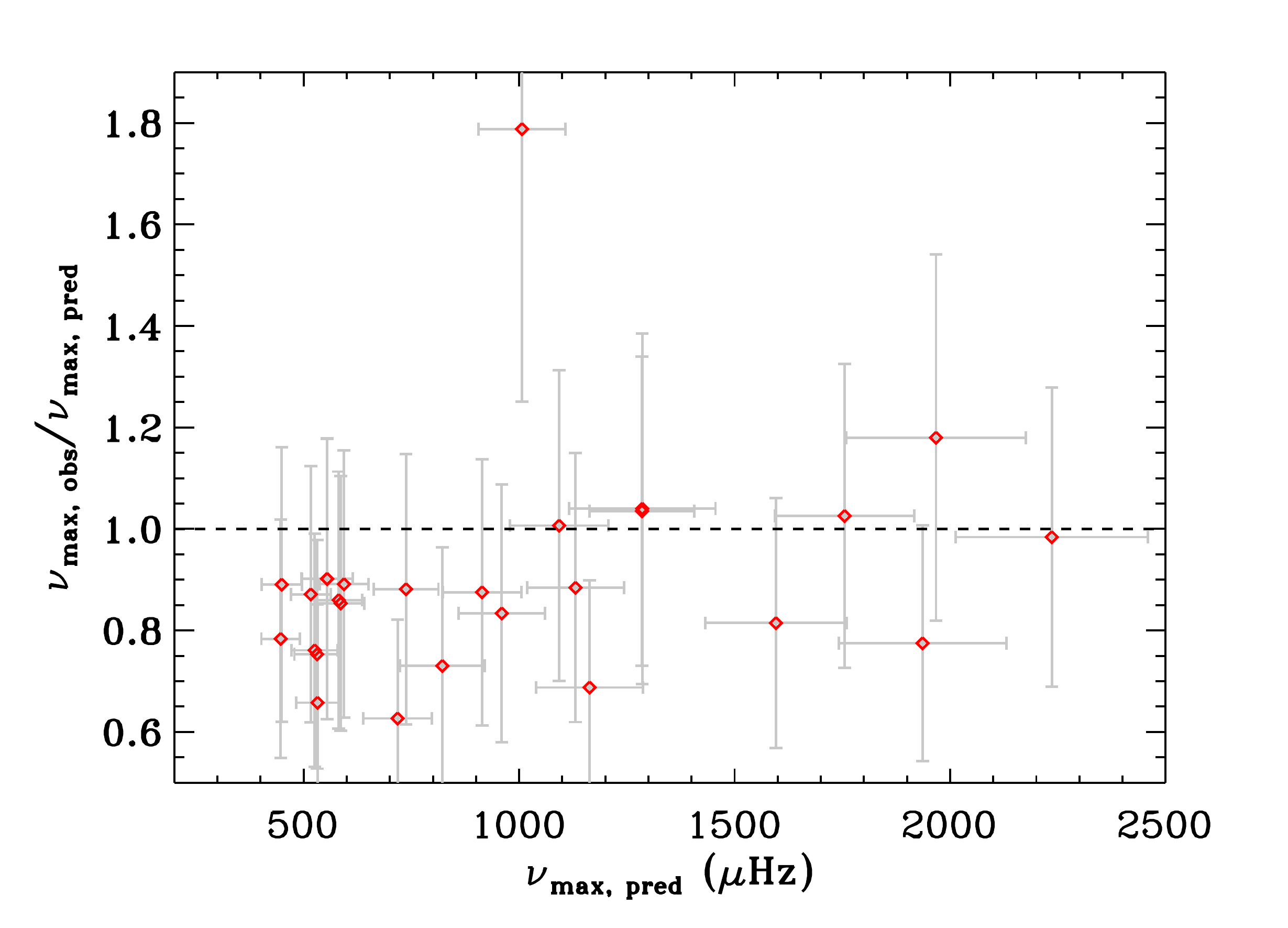}
\caption{Same as Figure~\ref{numax_comp} for the targets with possible detection of solar-like oscillations represented with red diamonds.}
\label{numax_comp_cand}
\end{center}
\end{figure}

\noindent During the visual checks a few interesting cases were flagged. KIC~2578869 seems to show modes only below $\nu_{\rm max}$, which is not something we have seen in the past and thus this star was put in the candidate list. KIC~11498538 has a strong rotation peak and the background would suggest that the modes are around 850\,$\mu$Hz, however the modes are barely visible probably due to the high level of activity of the star \citep[e.~g.][]{2019FrASS...6...46M,2020A&A...639A..63G}. The last star that was flagged is KIC~11818430, which seems to present two signals in the \'echelle diagram and might be a binary star.

\noindent Like for the confirmed sample, we searched for signature of rotation modulation in the lightcurves of these candidates. We find that 15 stars have a reliable $P_{\rm rot}$,  among which three stars are flagged as close binary candidates. The magnetic activity proxy values of the candidates are similar to the Sun from minimum to maximum magnetic activity (see Figure~\ref{Sph_Prot_cand}).


\begin{figure}[!h]
\begin{center}
\includegraphics[width=9cm]{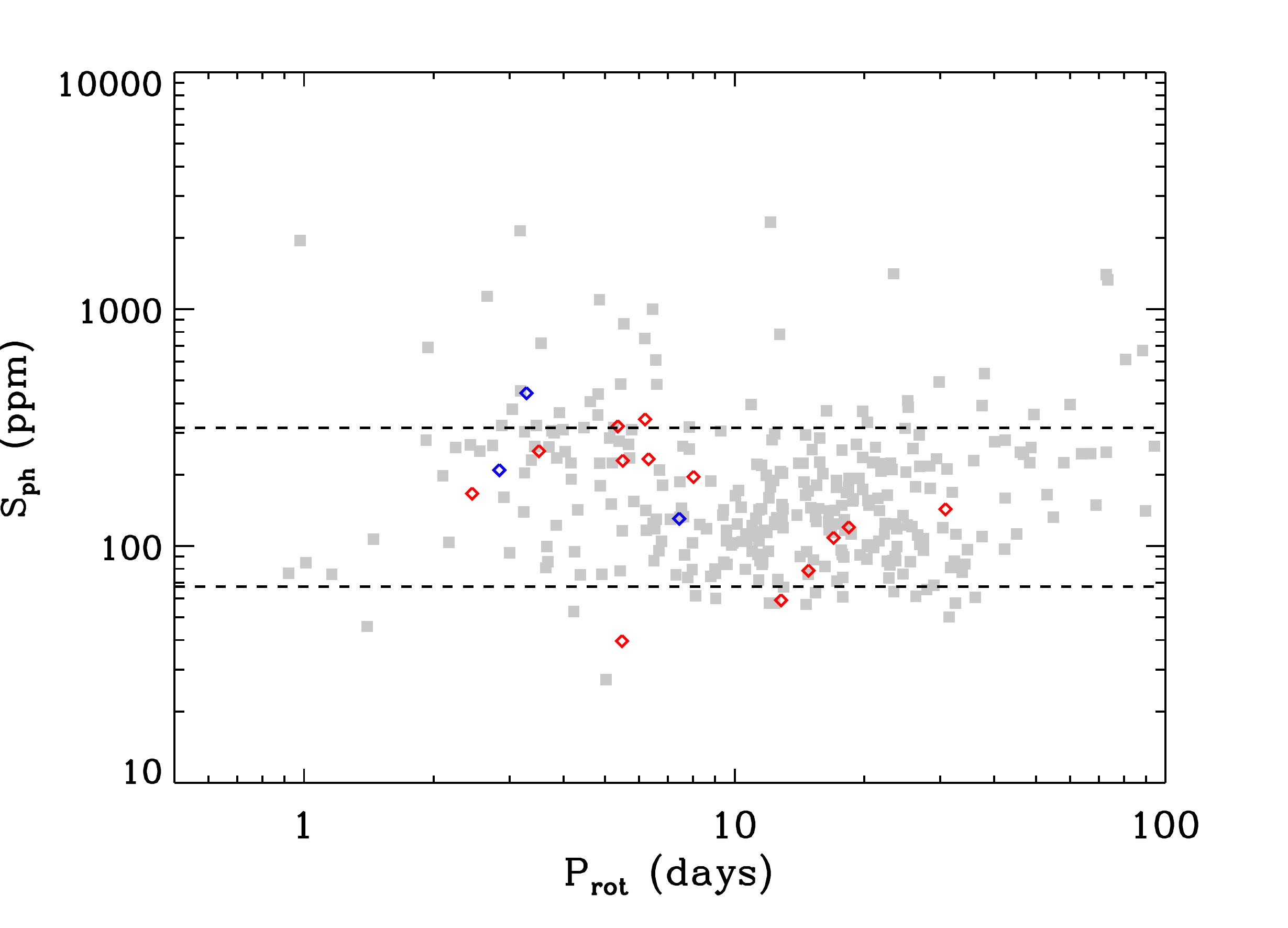}
\caption{Same as Figure~\ref{Sph_Prot} for the targets with possible detection of solar-like oscillations and with detection of rotation modulation.}
\label{Sph_Prot_cand}
\end{center}
\end{figure}

\noindent Finally, the magnitude distribution of the candidates, shown in Figure~\ref{Kp_histo_cand}, is similar to the one for the C14 sample, peaking at a magnitude of 11.

\begin{figure}[!h]
\begin{center}
\includegraphics[width=9cm]{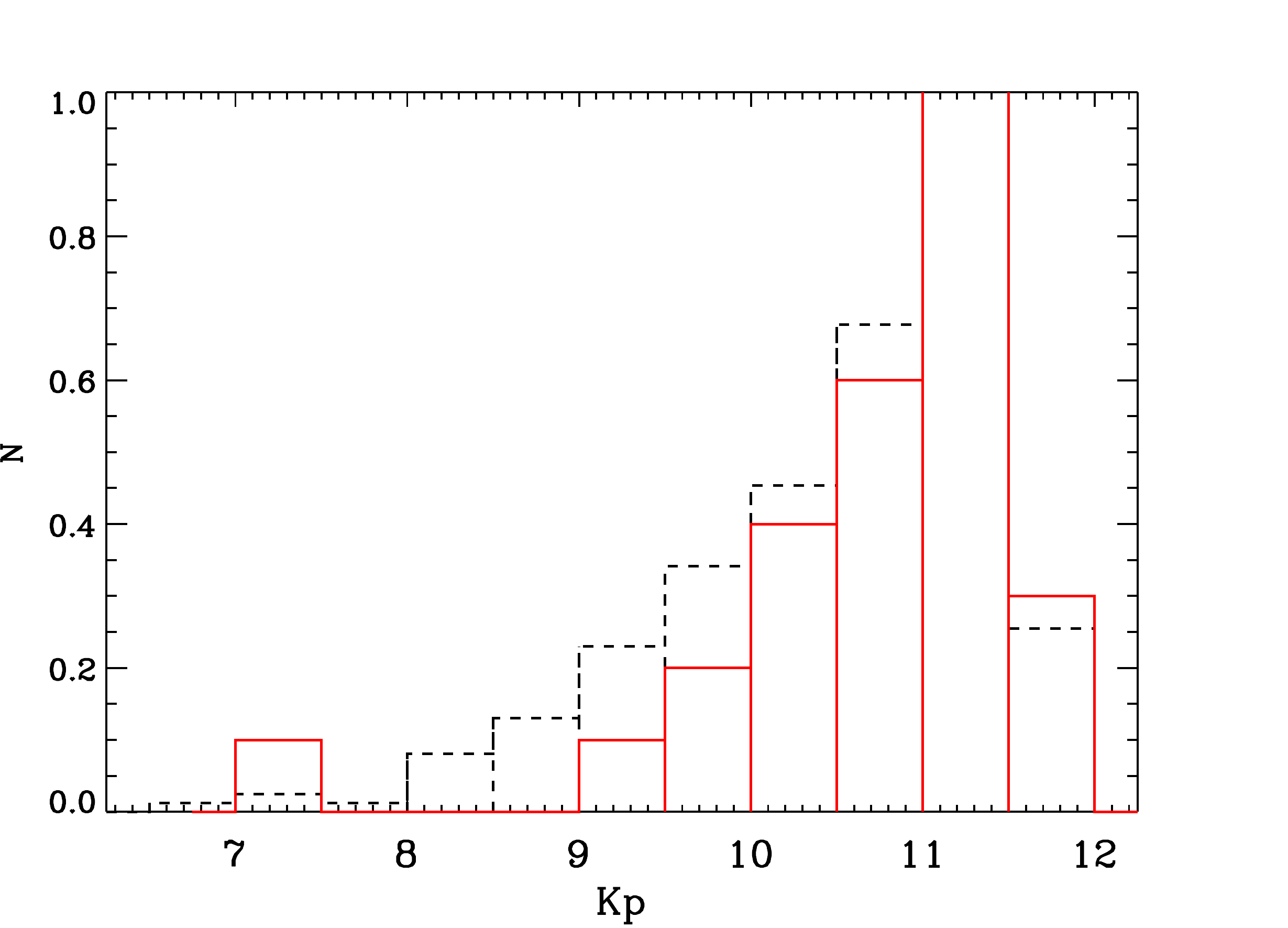}
\caption{Normalized histogram of the \kep\, magnitude for the C14 sample (black {  dashed} line) and the targets with possible detection of solar-like oscillations (red solid line). }
\label{Kp_histo_cand}
\end{center}
\end{figure}

\begin{table*}[h]
\caption{Seismic and stellar parameters of the 26 candidates with possible seismic detections. Frequency of maximum oscillation power, $\nu_{\rm max}$, is approximated. {  Same flags as in Table~\ref{tab1}.}}
\begin{center}
\begin{tabular}{cccccccc}
\hline
\hline
KIC & $T_{\rm eff}$ (K)& $\log g$ (dex)& [Fe/H] (dex)& $\nu_{\rm max}$ ($\mu$Hz)& Flag  \\
\hline
1570713 &  6814\,$\pm$\, 147& 3.95\,$\pm$\, 0.03&  0.01\,$\pm$\, 0.16& 1100&2\\
1571088 &  5957\,$\pm$\, 125& 3.86\,$\pm$\, 0.08& -0.11\,$\pm$\, 0.02&  500&0\\
2834481 &  6017\,$\pm$\, 128& 4.07\,$\pm$\, 0.08& -0.13\,$\pm$\, 0.01& 1000&0\\
3329439 &  6307\,$\pm$\, 138& 4.27\,$\pm$\, 0.07&  0.10\,$\pm$\, 0.01& 1500&0\\
3425564 &  6239\,$\pm$\, 136& 3.96\,$\pm$\, 0.08&  0.05\,$\pm$\, 0.01&  650&0\\
3633694 &  6531\,$\pm$\, 149& 3.89\,$\pm$\, 0.07&  0.08\,$\pm$\, 0.02&  450&0\\
3750375 &  6110\,$\pm$\, 129& 3.75\,$\pm$\, 0.08&  0.08\,$\pm$\, 0.02&  500&0\\
7601803 &  6213\,$\pm$\, 134& 3.67\,$\pm$\, 0.08&  0.09\,$\pm$\, 0.03&  400&0\\
7699517 &  6041\,$\pm$\, 137& 3.86\,$\pm$\, 0.08&  0.10\,$\pm$\, 0.01&  800&0\\
7770559 &  6640\,$\pm$\, 132& 4.06\,$\pm$\, 0.15& -0.08\,$\pm$\, 0.15& 1337&1\\
8587192 &  5791\,$\pm$\, 116& 4.18\,$\pm$\, 0.08&  0.19\,$\pm$\, 0.01& 1800&0\\
...\\
\hline
\end{tabular}
\end{center}
\label{tab4}
\end{table*}%

\begin{table*}[h]
\caption{Rotation periods and magnetic activity levels of the 15 stars with {  possible} seismic detections. }
\begin{center}
\begin{tabular}{ccc}
\hline
\hline
KIC &  $P_{\rm rot}$ (days) & $S_{\rm ph}$ (ppm)  \\
\hline
1430239 &   18.39\,$\pm$\,  3.04&  119.8\,$\pm$\,    4.4\\
1570713 &    5.49\,$\pm$\,  0.58&  228.9\,$\pm$\,   14.8\\
2834481 &   14.83\,$\pm$\,  1.82&   78.6\,$\pm$\,    3.9\\
3329439 &    5.47\,$\pm$\,  0.41&   39.6\,$\pm$\,    4.0\\
3425564 &    3.51\,$\pm$\,  0.63&  251.3\,$\pm$\,   18.9\\
3750375 &    5.35\,$\pm$\,  1.04&  320.0\,$\pm$\,   18.7\\
7601803 &    6.31\,$\pm$\,  1.11&  232.6\,$\pm$\,   13.6\\
7699517 &   16.95\,$\pm$\,  1.47&  108.2\,$\pm$\,    4.7\\
8604757 &    7.43\,$\pm$\,  0.52&  130.2\,$\pm$\,    7.8\\
8783286 &   30.84\,$\pm$\,  2.65&  142.9\,$\pm$\,    4.0\\
8947442 &    2.46\,$\pm$\,  0.17&  166.3\,$\pm$\,   15.3\\
8959788 &    8.02\,$\pm$\,  1.70&  195.6\,$\pm$\,   10.6\\
10014894 &   12.80\,$\pm$\,  1.00&   59.0\,$\pm$\,    3.8\\
11498538 &    3.28\,$\pm$\,  0.21&  441.6\,$\pm$\,   33.4\\
11818430 &    6.19\,$\pm$\,  0.82&  342.2\,$\pm$\,   19.4\\

\hline

\end{tabular}
\end{center}
\label{tab4}
\end{table*}%

\newpage
\section{Rotation analysis examples}

{  We show here some examples of lightcurves of stars with measured  rotation periods reported in this paper along with the rotation analysis as described in Section~\ref{sec:rot}.}

\begin{figure}[!h]
\begin{center}
\includegraphics[width=10cm]{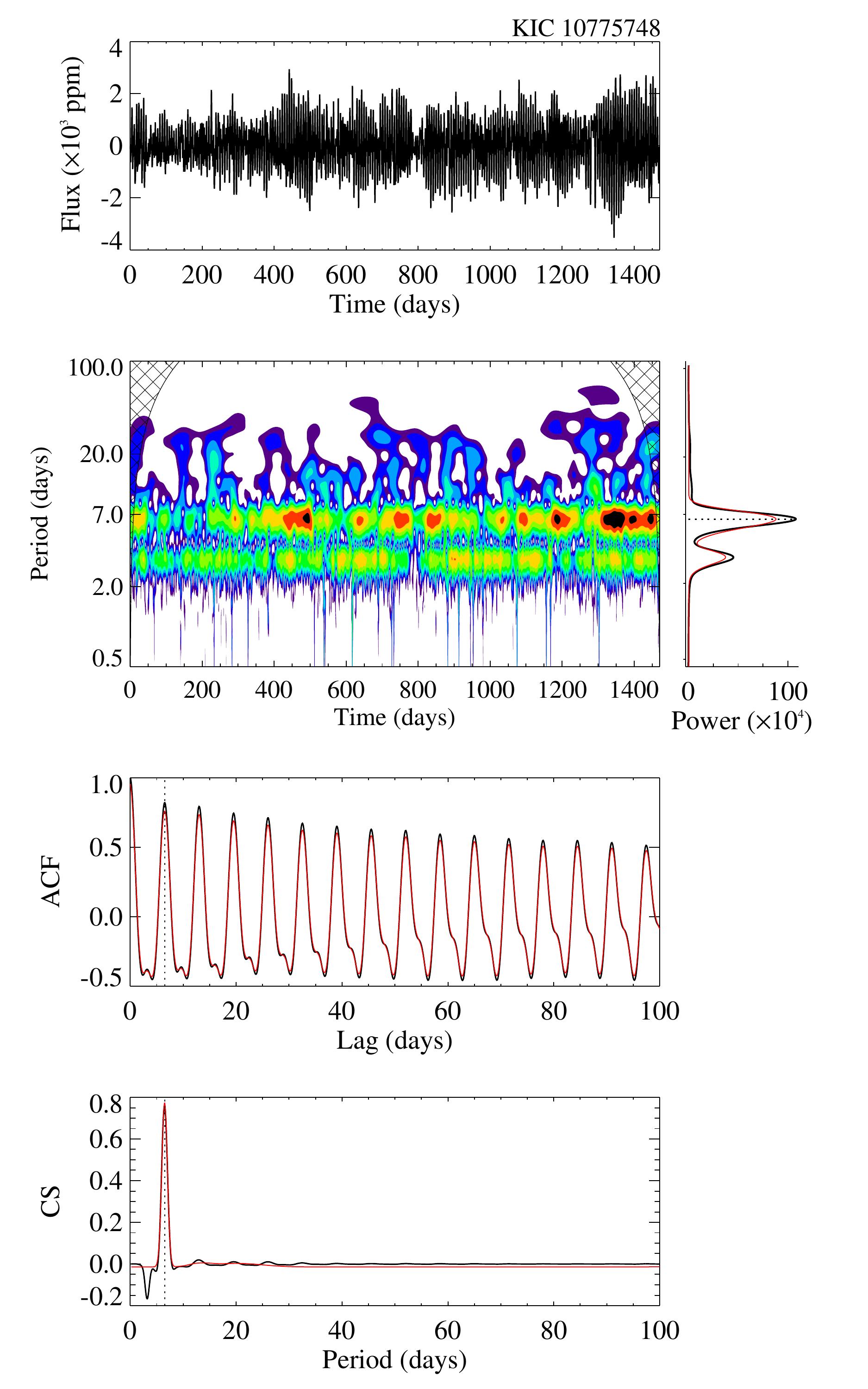}
\caption{{  Top row: KEPSEISMIC lightcurve for KIC~10775748. Second row: Wavelet power spectrum (left), and global wavelet power spectrum (right; black), where the best fit with multiple Gaussian functions is shown in red. Third row: Autocorrelation function (ACF) of the lightcurve in black and its smoothed version in red. Bottom row: Composite spectrum (black) and best fit with multiple Gaussian functions (red). The dashed lines mark the rotation-period estimates from each diagnostic.}}
\label{Rot1}
\end{center}
\end{figure}

\begin{figure}[!h]
\begin{center}
\includegraphics[width=10cm]{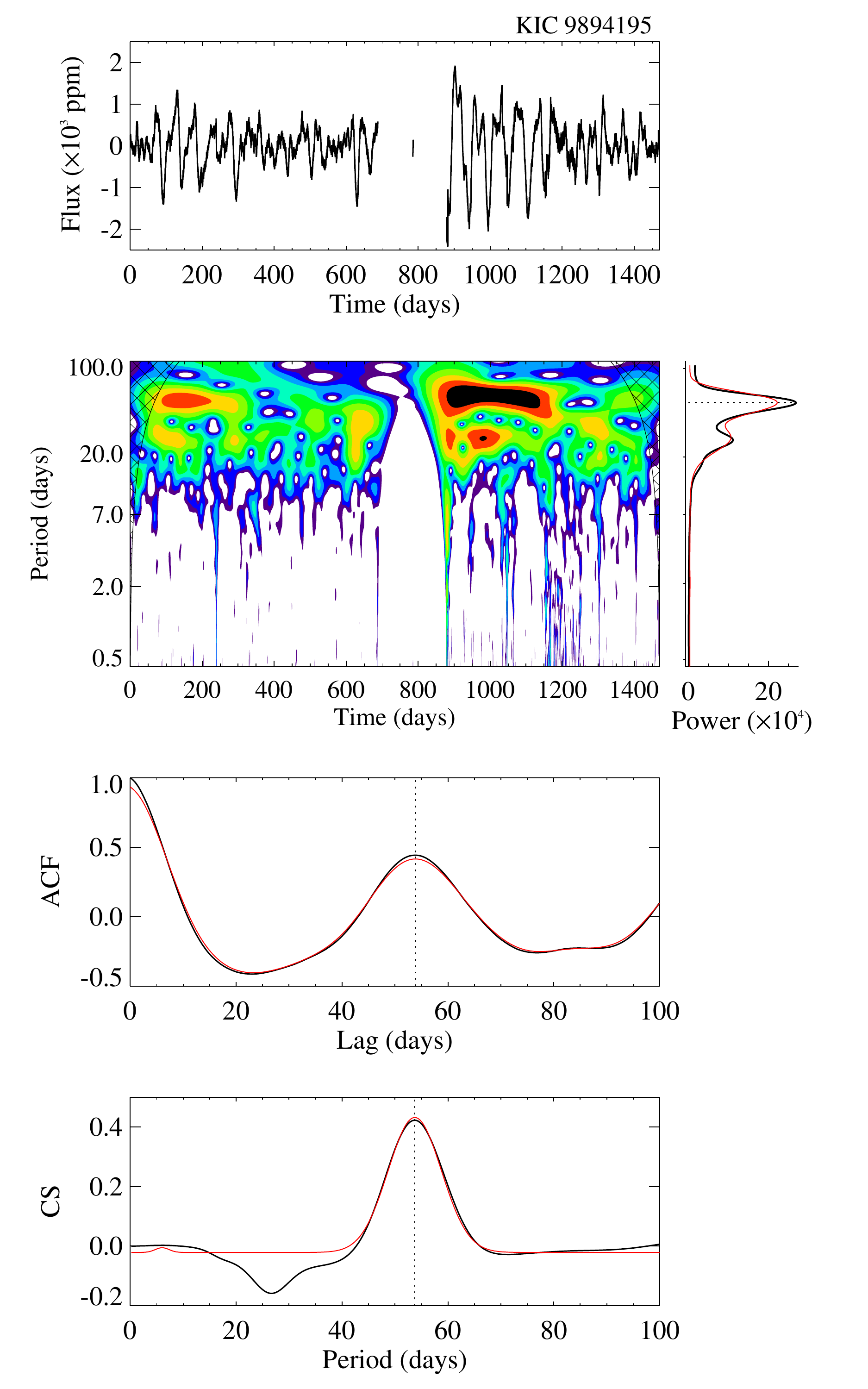}
\caption{Same as Figure~\ref{Rot1} for KIC~9894195.}
\label{Rot2}
\end{center}
\end{figure}

\begin{figure}[!h]
\begin{center}
\includegraphics[width=10cm]{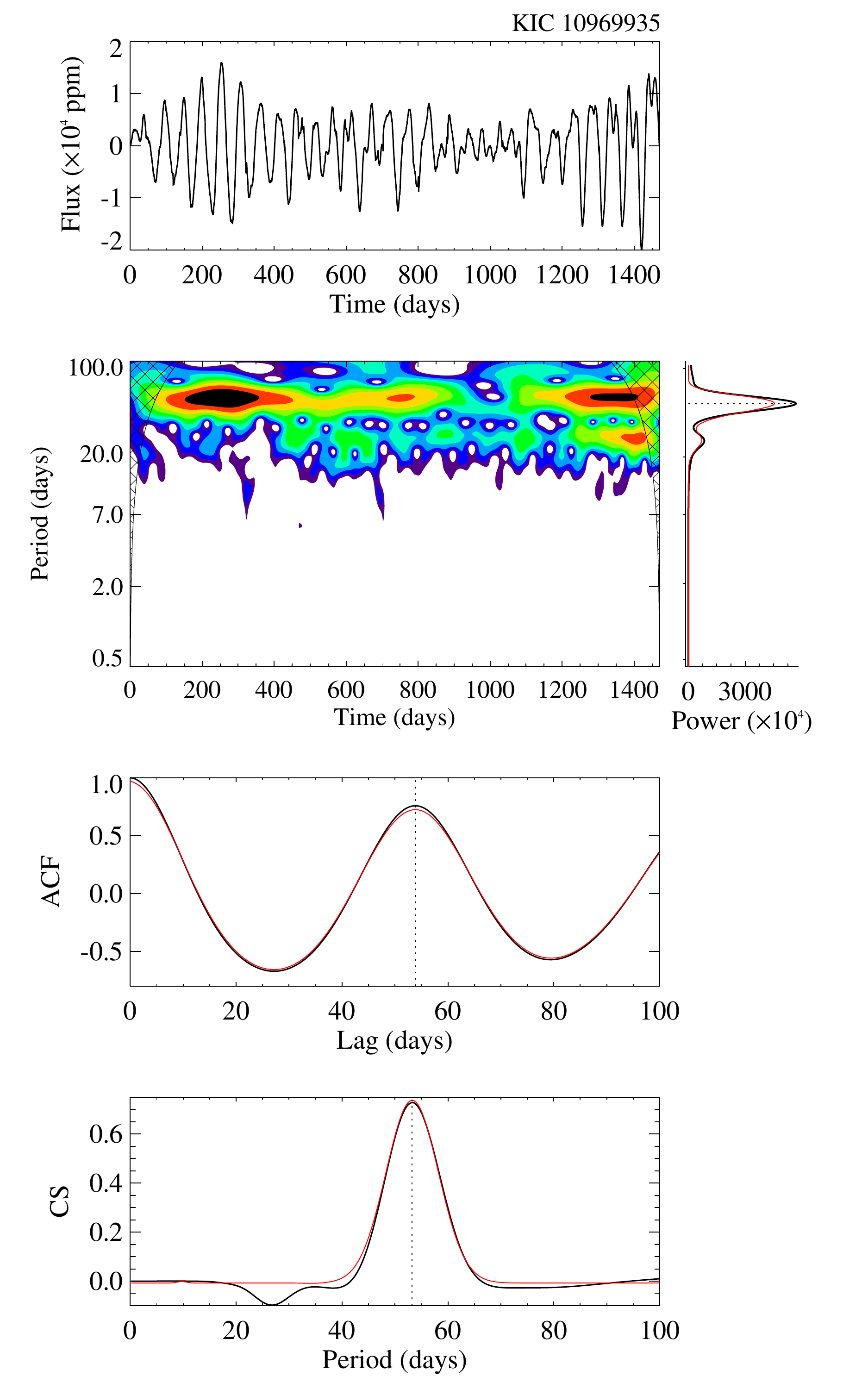}
\caption{Same as Figure~\ref{Rot1} for KIC~10969935.}
\label{Rot3}
\end{center}
\end{figure}

\newpage
\section{Radii comparison with a different $T_{\rm eff}$ scale}

We computed the seismic radii using the effective temperatures from B20. The comparison with the B20 radii is shown in Figure~\ref{RGaia_B20}. The agreement is slightly better than when using the spectroscopic $T_{\rm eff}$. 

\begin{figure}[!h]
\begin{center}
\includegraphics[width=9cm]{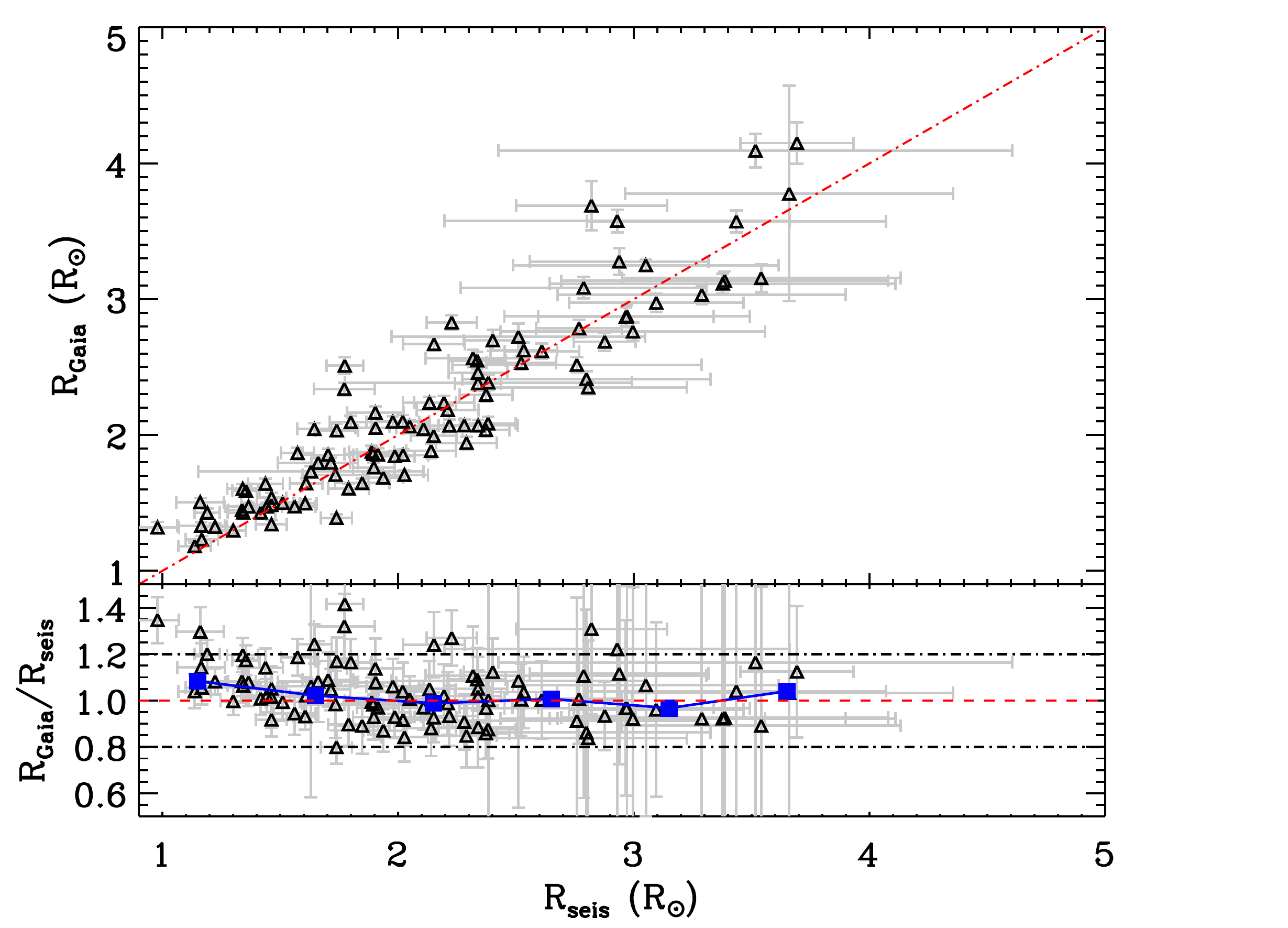}
\caption{Same as Figure~\ref{RGaia} but using the effective temperature from B20 in the seismic scaling relations to compute the radii, $R_{\rm seis}$.}
\label{RGaia_B20}
\end{center}
\end{figure}

\section{Full table of global seismic parameters for the 625 solar-like stars}

\begin{table*}[h]
\caption{{  Global seismic parameters from A2Z for the 625 stars detection of solar-like oscillations.}}
\begin{center}
\begin{tabular}{ccc}
\hline
\hline
KIC &  $\nu_{\rm max}$ ($\mu$Hz)& $\Delta \nu$ ($\mu$Hz)  \\
\hline
1430163 &   1807\,$\pm$\,   43& 85.71\,$\pm$\,  1.79\\
1435467 &   1369\,$\pm$\,   56& 70.80\,$\pm$\,  1.47\\
1725815 &   1040\,$\pm$\,   28& 55.97\,$\pm$\,  1.43\\
2010607 &    675\,$\pm$\,    7& 42.44\,$\pm$\,  1.49\\
2010835 &   1312\,$\pm$\,   19& 72.73\,$\pm$\,  4.52\\
2309595 &    646\,$\pm$\,   16& 38.97\,$\pm$\,  1.38\\
2450729 &   1078\,$\pm$\,   36& 61.05\,$\pm$\,  1.95\\
2578869 &    809\,$\pm$\,   26& 48.87\,$\pm$\,  1.89\\
2837475 &   1638\,$\pm$\,   72& 75.71\,$\pm$\,  1.22\\
2849125 &    729\,$\pm$\,   31& 40.44\,$\pm$\,  1.29\\
2852862 &    988\,$\pm$\,   61& 54.68\,$\pm$\,  1.07\\
2865774 &   1260\,$\pm$\,   37& 64.20\,$\pm$\,  2.16\\
2991448 &   1127\,$\pm$\,   34& 61.22\,$\pm$\,  2.91\\
2998253 &   2034\,$\pm$\,    9& 89.00\,$\pm$\,  2.13\\
3102595 &   1071\,$\pm$\,   17& 52.20\,$\pm$\,  1.66\\
3112152 &   1263\,$\pm$\,   43& 65.00\,$\pm$\,  1.77\\
3112889 &    817\,$\pm$\,   30& 53.09\,$\pm$\,  1.86\\

...\\

\hline

\end{tabular}
\end{center}
\label{tabE}
\end{table*}

\end{document}